\documentclass[useAMS,usenatbib]{mn2e}

\usepackage{amsmath,graphicx,wrapfig,lipsum,caption,overpic,amssymb}
\usepackage{subfigure}

\newcommand{\ignore}[1]{}

\title[Radial Drift of Dust in Protoplanetary Disks]{Radial Drift of Dust in Protoplanetary Disks: The Evolution of Ice lines and Dead zones}
\author[A. J. Cridland, Ralph E. Pudritz \& T. Birnstiel]{A. J. Cridland$^{1}$\thanks{E-mail:
cridlaaj@mcmaster.ca}, Ralph E. Pudritz$^{1,2,3,4}$\thanks{E-mail:
pudritz@mcmaster.ca} and T. Birnstiel$^4$\\
$^{1}$Department of Physics and Astronomy, McMaster University, Hamilton, Ontario, Canada, L8S 4E8 \\ $^2$Origins Institute, McMaster University, Hamilton, Ontario, Canada, L8S 438 \\ $^3$Zentrum f\"ur Astronomie der Universit\"at Heidelberg, Institut f\"ur Theoretische Astrophysik, Albert-Ueberle-Str. 2, 69120 Heidelberg, Germany \\ $^4$Max Planck Institut f\"ur Astronomie K\"onigstuhl 17, D-69117 Heidelberg, Germany}

\begin{document}

\date{\today}

\pagerange{\pageref{firstpage}--\pageref{lastpage}} \pubyear{2015}

\maketitle

\label{firstpage}

\begin{abstract}
We have developed a new model for the astrochemical structure of a viscously evolving protoplanetary disk that couples an analytic description of the disk's temperature and density profile, chemical evolution, and an evolving dust distribution. We compute evolving radial distributions for a range of dust grain sizes, which depend on coagulation, fragmentation and radial drift processes. In particular we find that the water ice line plays an important role in shaping the radial distribution of the maximum grain size because ice coated grains are significantly less susceptible to fragmentation than their dry counterparts. This in turn has important effects on disk ionization and therefore on the location of dead zones. In comparison to a simple constant gas-to-dust ratio model for the dust as an example, we find that the new model predicts an outer dead zone edge that moves in by a factor of about 3 at 1 Myr (to 5 AU) and by a factor of about 14 by 3 Myr (to 0.5 AU). We show that the changing position of the dead zone and heat transition traps have important implications for the formation and trapping of planets in protoplanetary disks. Finally, we consider our results in light of recent ALMA observations of HL Tau and TW Hya.
\end{abstract}


\section{ Introduction }\label{sec:intro}

Dust plays a key role in the physics of protoplanetary disks, and hence the physics of planet formation. The largest grains most strongly contribute to the disk's opacity while the smallest grains provide the most surface by mass for the freeze out of volatiles. So understanding the evolution of the grain sizes and dust surface density will have an impact on the abundance of volatiles and free electrons in the disk, and hence our understanding of planet formation and the accretion of planetary atmospheres.

In the absence of small scale instabilities (for example: streaming instability \citep{YG05,R15} and zonal flows \citep{J11}) the dust grain sizes and surface density evolve through coagulation \citep{DD05}, fragmentation \citep{BW00}, and radial drift \citep{Whip73,W77}. 

These are competing processes, with rates that depend on the size of the grain and the density of the surrounding gas. It has been shown (eg. in \cite{B12}) that these processes can be modeled by simple analytic functions, and their relative relevance can be determined by comparing their physical timescales. This has been implemented in the `two-population dust' model of \cite{B12} and used to reproduce the observed properties of protoplanetary disks (ie \cite{B15}).

The purpose of this work is to combine an evolving astrochemical model of protoplanetary disks with a model for an evolving dust mass distribution. We show that a more complete model of dust physics which includes coagulation, fragmentation and radial drift, produces an ionization structure which evolves in a different manner than in a fiducial dust model with a constant dust-to-gas ratio. This new ionization structure has implications on the size and evolution of the disk's dead zone and ultimately on the formation history of planets forming within the disk. For instance in determining the radial migration rate of a forming planet, the net torque acting on the planet depends on its location relative to the edge of the dead zone.

The relevant background information is presented in \S \ref{sec:background}. In \S \ref{sec:method} the `two-population dust' (Two-pop-dust) model of \cite{B12} (or see \cite{B16}) and the astrochemical disk model from \cite{Crid16} are presented. In \S \ref{sec:results} we will demonstrate that the ice line causes an amplification of dust surface density within its location and that this new dust surface density distribution changes how the dead zone evolves as the disk ages. We will also compare this new evolution and structure of the dead zone with the structure of the dead zone from the simpler model. In \S \ref{sec:dis} we discuss the implications that the new model on planet formation and present our conclusions.

\section{ Background }\label{sec:background}
\subsection{ Dust, ionization, and the impact on planet formation }

The grain size and surface density distribution of the dust affects the ionization structure of the disk. In the Rayleigh limit of Mie theory, the efficiency of absorption and scattering of radiation scales as a power law with the size of the dust grain \citep{Tie05}. So the amount of ionizing radiation that penetrates into the disk depends on the rate of dust coagulation versus dust fragmentation. Radial drift plays an important role in the retention of these large grains because the sub-Keplarian orbits that are imposed on the largest grains by gas drag accretes these grains faster than viscous evolution alone \citep{W77,Br08,B10}. We expect that the radial structure of the disk's ionization will be sensitive to the concentration of large dust grains that are present through the disk, hence on the efficiency of radial drift. Fragmentation impacts this efficiency by limiting the maximum size of the grains at different radii.

Generally there is an inner region of the disk that is dominated by thermal ionization, an intermediate region where the ionization is low,  and an outer region of radiation dominated ionization at larger radii \citep{Gammie96,MP03}. We assume that disk turbulence is produced through the magnetorotational instability (MRI, \cite{BH94}), which relies on a coupling of the disk's magnetic field with the electrons in the gas. This region of low ionization leads to weak MRI driven turbulence and is called the `dead zone' \citep{Gammie96,MP03,Dutrey14}. The point where one region transitions to the other - the `edge' of the dead zone - has important implications on planet formation via the combination of the core accretion \citep{IL04} and `planet trap' models \citep{M06,HP10,LM12,HP13,Crid16}.

The planet trap model of planet migration focuses on the fact that there are inhomogeneities in the physical properties of accretion disks (ie. temperature, opacity) that slow the rate of planetary migration via Type I migration \citep{M06,HP13}. We assume that a planet that is trapped will always be located at the radial location of the planet trap until the planet has grown massive enough to open a gap and begin Type-II migration. The edge of the dead zone acts as one of these planet traps \citep{MP03}, so accurately modeling the dust physics has implications on where a growing planet will form.

\ignore{
Additionally, \cite{Crid16} demonstrated that the relative location of the dead zone and the other planet traps affected whether a planet will remain trapped for the duration of its formation. Planet trapping occurs because of the relative strength and direction of the Lindblad and co-rotation torques. Turbulence is required for the co-rotation torque to contribute a net torque on the planet \citep{Exo10,Baru14,Cole14}, otherwise angular momentum transport is dominated by the Lindblad torque and trapping is unlikely.
}

The number of small grains also impact the ionization structure of the disk because they offer the highest surface area per mass of dust for the freeze out of volatiles, and the capture of electrons. If fragmentation was inefficient, the number of small grains is quickly reduced \citep{DD05}, eliminating an important sink for electrons. In our work, we assume that the grains are either singly charged, or neutral. A dust grain capturing a single electron is a good assumption for the smallest grains \citep{Ak15} but becomes a worse assumption for larger grains because they can efficiently capture multiple electrons \citep{Ak14}. Under this assumption as the average grain size grows, the total surface area available for electron capture is reduced, causing the ionization to increase throughout the disk. It is currently unclear what the effect of multiple electron capture has on the global ionization structure of the disk, and this is beyond the scope of this work.

\subsection{ Limiting the dust grain size }

The maximum dust grain size at a given orbital radius is limited by two processes: radial drift \citep{Whip73,W77} and fragmentation \citep{BW00,DD05}. The maximum grain size of a population is set by radial drift when the drift rate of a large grain $a_{max}$ is equal to its growth rate from the population of smaller grains. Any grain larger than $a_{max}$ will accrete onto the host star faster than it can be built from the coagulation of smaller grains \citep{B12}. Similarly, the maximum size is limited by fragmentation when the fragmentation rate of a large grain is equal to its growth rate. Globally, the size distribution of the dust in protoplanetary disks is set by these barriers, and their impact has begun to be observed by the Atacama Large Millimeter Array (ALMA) \citep{CP12,ALMA15,Z15,And16,Taz16}.

The shape of the dust size distribution can be described by a broken power law (see below), whose functional form has been motivated by observations \citep{MRN77} and reproduced by numerical simulations \citep{DD05,Br08,B12}. This power law extends from the smallest grain of approximately 0.1 microns up to a maximum grain size that is set by the physical barriers described above. In protoplanetary disks, it has been shown that fragmentation and coagulation produces a maximum grain that varies with disk radius as a broken power law \citep{B11}.

A difference between the two barriers is how they impact the surface density of the dust. The fragmentation of the largest dust grains does not remove any dust mass from the disk, it simply redistributes the dust surface density from the largest grains to the smaller ones \citep{DD05}. Conversely radial drift does reduce the surface density of dust, by forcing the largest dust grains into sub-Keplerian orbits which spiral into the host star \citep{W77}. This difference can be seen when comparing the evolution of the surface density distribution of the dust in simulations that do not have radial drift \citep{DD05} with those that do \citep{Br08}.

\ignore{
It was pointed out in \cite{B09} that the retention of dust is dependent on the grain's fragmentation threshold speed. This speed describes the minimum relative speed of two colliding grains that will lead to fragmentation. A higher threshold speed leads to larger dust grains because the grains are less likely to fragment during a collision. The threshold speed is dependent on the composition and structure of the dust grain, as well as the quantity of ice that is frozen onto the dust grain. A grain that is fully covered in water ice will be less susceptible to fragmentation than grains that contain no ice coverage \citep{Wa09,Gund15}.
}

The efficiency of grain fragmentation is dependent on the layer of ice that forms on the grain outside the water `ice line' of the disk \citep{B09,B10,Ban15}. The water ice line defines the location in the disk where the water content transitions between being primarily in the vapour phase, to being primarily in the ice phase. At radii outside the location of the ice line the grains will be strengthened by their ice content, while within the ice line the grains are weaker since the water content is in the vapour phase. 

The dry grains will generally be weaker and hence susceptible to fragmentation compared with grains outside the ice line, and hence they will radially drift slower. This less efficient radial drift means that there will be an enhancement of dust surface density within the ice line because it will not be cleared out as fast as it would if the fragmentation rate was constant \citep{B10}.

\subsection{ Linking dust grain size and astrochemistry }

In what follows, the global structure of the ionization will be shown to depend on the dust, through physical processes like fragmentation, radial drift and coagulation. The effect of grain size on astrochemical networks is through the availability of dust grains on which gas species freeze, as well as the availability of grains to capture electrons; thereby reducing the disk's ionization. Numerically, its effect is parameterized with a code parameter known as the `freezing efficiency' which depends on the average size of the dust grains as computed by our dust model. Hence the freezing efficiency will have a radial dependence on the number and size distribution of dust grains.


\section{ Method }\label{sec:method}

Generally, our method is as follows: first we produce the gas surface density and temperature radial distribution based on the analytic model of \cite{Cham09}. This model assumes that the gas temperature can be described by three power laws depending on the heating source and temperature dependence of the disk's opacity. In the inner region (T $>$ 1380 K) of the disk where the dust sublimates, the disk opacity varies as a power law in temperature. At lower temperatures we assume that the opacity is constant. The disk is heated by two sources: viscous heating caused by the accretion process, and direct irradiation from the host star.

Second we compute the surface density radial distribution of the dust based on either of our two dust models below. The CPA16-dust model assumes that gas and dust are perfectly mixed with a constant gas-to-dust ratio, while the Two-pop-dust model computes the coagulation, fragmentation, and radial drift of dust grains. Both dust models assume that the gas evolves in the same way, set by the gas model presented in \cite{Crid16}. In principle the gas evolution will depend on the evolution of the dust, however we are primarily interested in how the ionization state of the gas depends on the dust model. And hence we assume that the gas evolution is not affected by changes in the radial and size distribution of the dust grains.

Next we compute the UV and X-ray radiation field in the disk using RADMC3D and the results from the gas and dust models. This Monte Carlo radiative transfer scheme use $10^7$ photon packets to compute the radiation field. The flux of this ionizing radiation is sensitive to the choice of opacity tables, we use the optical constants of \cite{Draine03}, and compute the opacity using the on-the-fly method of RADMC3D. 

Finally, we compute the chemical distribution with the radiation field, gas and dust volume density and temperature as inputs. The chemical structure is computed over 300 `snapshots' of the disk across 4.1 million years of disk evolution. Each snapshot represents the disk at individual times throughout the disk's life time, evenly space between 0.1 - 4.1 million years. Of particular importance to this work is the distribution of free electrons, as they are responsible for the coupling to the magnetic field that leads to turbulence.

\subsection{ CPA16-dust Model }\label{sec:chamdust}

In \cite{Crid16} (CPA16) we used the analytic disk model of \cite{Cham09} and a constant dust-to-gas ratio of 0.01 to describe the radial distribution of the dust. Built into this model (CPA16-dust) is the assumption that the dust surface density perfectly traces the surface density of the gas and the two fluids are fully mixed.

The \cite{Cham09} disk model is a self-similar solution to the diffusion equation, which results in a power law for the surface density of gas and dust: $\Sigma \propto R^s$. The value of the exponent $s$ is determined by the heating details, which results in three regions: evaporative, viscous and radiative. At temperatures below the evaporation temperature of silicate grains ($\sim 1380$ K) we assume that the average opacity is constant in the disk. At higher temperatures, in the evaporative region, the opacity varies with temperature as a power law. The evaporative and viscous regions are heated through the viscous stresses in the gas caused by the disk's mass accretion while the radiative region is heated through the direct illumination from the host star. For a more detailed discussion of our disk model see CPA16 or \cite{APC16}.

A constant opacity through the disk is an over simplified assumption considering the dust model below. However more complicated analytic models (ie. \cite{St98}), which include a temperature dependence of the opacity show minimal deviations in temperature and surface density profiles when compared to the model presented in CPA16. Additionally, these models ignore the evolution of the size distribution and surface density of dust as the disk ages. A full analytic description of the opacity which includes the effect of dust evolution has yet to be developed, and we require an easily computed disk model to combine the dust evolution, radiative transfer and astrochemistry.

\subsection{ Two-pop-dust Model }\label{sec:tpopd}

To compute the dust evolution with the Two-pop-dust model we modified the algorithm that was developed by \cite{B12} so that the gas component evolved according to the analytic model of \cite{Cham09}. This allows us to directly compare the resulting ionization structure due to the variations in dust evolution between the two models, rather than the gas evolution.

A key component of the Two-pop-dust model is determining the size distribution of the dust grains. As mentioned before, the maximum grain size is set by an equilibrium between either fragmentation or radial drift with coagulation. Once the maximum grain size is determined, the size distribution up to the maximum size can be described reasonably well by a power law \begin{align}
n(m)dm &= Am^{-\alpha}dm,
\label{eq:01}
\end{align}
where $A$ and $\alpha$ are positive constants \citep{D69,Tan96,Mak98}. These constants depend on whether fragmentation and radial drift dominates the evolution of the largest grains \citep{B11}. In the region dominated by fragmentation the maximum grain size is given by \citep{B12}: \begin{align}
a_{max} = a_{frag} \propto \frac{\Sigma_g}{\rho_s\alpha}\frac{u_f^2}{c_s^2},
\label{eq:02}
\end{align}
where $\Sigma_g$ is the surface area of the gas, $\rho_s$ is the average density of the dust grains, $\alpha$ is the turbulent parameter \citep{SS73}, $c_s$ is the gas sound speed and $u_f$ is the collision speed above which particle collisions lead to fragmentation \citep{B12}.

The second constraint on the maximum size of the grain is the rate of radial drift. \cite{B12} showed that the maximum grain size in the radial drift dominated regime is given by: \begin{align}
a_{max} = a_{drift} \propto \frac{\Sigma_d}{\rho_s}\frac{V_k^2}{c^2_s}\gamma^{-1}
\label{eq:03}
\end{align}
where $\Sigma_d$ is the surface density of the dust, $V_k$ is the Keplarian velocity and \begin{align}
\gamma = \left|\frac{d \ln P}{d \ln r}\right|,
\label{eq:04}
\end{align}
is the absolute value of the power-law index of the gas pressure profile \citep{B12}.

The Two-pop-dust model has been calibrated to the numerical result of \cite{B10}, which included coagulation, fragmentation and radial drift. This simpler model computes the coagulation and fragmentation processes between only two populations of grains, one small (monomer) size and one larger size. Once the total surface density is computed, the size distribution of the grains is reconstructed based on the result of \cite{B15}. This model benefits from an increase in computational efficiency because only two sizes are resolved, and well reproduces the dust surface density evolution. 

In choosing a value for $u_f$ we follow the work of \cite{B10}. Within the ice line the grains are not covered with ice, and have a fragmentation threshold speed of 1 m/s \citep{BW08}. Outside of the ice line the grains are strengthened by a layer of ice, resulting in a fragmentation threshold speed of 10 m/s \citep{Gund16}. 

In CPA16 the ice line was defined as the disk radius at which half the water vapour had been converted to water ice. However the ice line is in fact a region of space with a width of a few tenths of AU (see CPA16, their Figure 6). Within this transition region the grains will not be perfectly covered in ice, and we assume that the covered fraction of the grain corresponds with the ratio of water ice at any time to the final water ice abundance. Additionally we assume that the grain is strengthened by an increase in the ice coverage, so that the fragmentation threshold speed varies smoothly with the ice coverage of the grains (see Equation \ref{eq:meth01}).

This assumption ignores the effects of sintered grains. Sintering is the process of sticking grain aggregates together at temperatures below the sublimation temperature of specific volatiles. Sintered grains are characterized by frozen `necks' that connect the aggregates (eg. \cite{Sirono11}). These necks can actually {\it weaken} the grains during collisions because they cannot dissipate collisional energy into deformation. Instead, sintered grains will tend to fragment at lower collisional speeds than their unsintered counterparts. It has been suggested that sintering regions around the sublimation point of a few volatiles can explain the dust emission observed in HL Tau \citep{Oku16}. In their work, there is no difference between the fragmentation threshold speed in the inner regions of the disk and the regions outside of the ice line. Instead, the fragmentation threshold speed is modified within the sintering regions of every major volatile in the disk. Additionally, around the edge of the ice line there 
is little qualitative difference between the dust distribution in a sintering model and the distribution we will compute in this work.

Our choice of fragmentation threshold speed differs from the work of \cite{Oku16}. In their work the threshold speed is 20 m/s and 50 m/s in the sintered and unsintered regions respectively. The unsintered region would be equivalent to our ice-covered region where $u_f\sim 10$ m/s, while their sintered region would be equivalent to the transition region where $u_f$ is a function of radius. These higher threshold speeds are generally attributed to sub-micron sized grains and are based on the numerical results of \cite{Saf07} and \cite{Wa09}. As the grain grows, lab experiments have shown that the threshold speed decreases with increasing grain size \citep{Bei11} because the strength and porosity of the grain becomes more important. We follow \cite{B12} in choosing a threshold speed of 10 m/s to act as a global average for the majority of the dust that is covered in ice.

The smooth transition between low and high threshold speeds is modeled with an arctan function of the form: \begin{align}
u_f(r) = a + b \arctan\{c \cdot(r-r_0)\}/\pi.
\label{eq:meth01}
\end{align}
The parameter	 $r_0$ is the center of the arctan function and roughly matches the location of the water ice line as defined in our previous work. The three other constants ($a$, $b$ and $c$) are fit so that $u_f(r << r_0) \sim 1$ m/s, $u_f(r >> r_0) \sim 10$ m/s, and the function transitions at the same rate as water transitions between the vapour and ice phases. 

We fit Equation \ref{eq:meth01} so that its slope is nearly the same as the slope of the water ice radial distribution from our fiducial disk model in CPA16 and find that the fitting parameters do not vary between the three times. The only parameter that varies in time is the radial location of the middle of the function ($r_0$). This is not surprising since the location of the ice line evolves as the disk ages, moving inward as the disk cools. We find that the position of $r_0$ traces a disk temperature of approximately 140 K which is just below the sublimation temperature of water. For this reason we define $r_0 \equiv r(T = 140 ~{\rm K})$. For the fiducial disk we find that the fitting constants have average values of $a = 5.47$ m/s, $b = 9.11$ m/s, and $c = 15$. These fitted values result in the limits $u_f(r >> r_0) = 10.025$ m/s and $u_f(r << r_0) = 0.915$ m/s.

\subsection{ Radiation Field }

We use RADMC3D \citep{RADMC} to compute the local X-ray and UV fields for our astrochemical model. RADMC3D is a Monte Carlo radiative transfer scheme that can handle the presence of multiple dust populations, by computing the wavelength dependent opacities of each grain population on the fly. In the CPA16-dust model we assume that the dust and gas are well mixed so that the scale height of the dust is the same as the gas.  For the absorption and scattering opacities we use the pre-computed values from \cite{WD01} and \cite{BB11}. In the Two-pop-dust model we used the on-the-fly method of calculating the dust opacities using the real and complex indices of refraction computed by \cite{Draine03} and assumed Mie scattering. We sampled the dust distribution produced with the \cite{B12} code with 20 different grain sizes equally spaced in log-space. Their surface densities are normalized so that the total dust surface density remained the same after the sampling. We converted the dust surface densities to volume 
densities (the preferred input of RADMC3D) by assuming an equilibrium between vertical turbulent mixing and gravitational settling. This assumption results in a simple conversion where \citep{Du95,HP10}:\begin{align}
\rho_{d}(a,r,z) &= \frac{\Sigma_{d}(a,r)}{\sqrt{2\pi} H_{d}(a,r)}\exp\left\{-\frac{z^2}{2H_{d}^2}\right\}.
\label{eq:meth02}
\end{align}
$\Sigma_{d}$ is the surface density outputted from the \cite{B12} code as a function of dust radius $a$ and orbital radius $r$. The scale height of the dust ($H_d$) also depends on the size of the dust grain and position in the disk:\begin{align}
\frac{H_d(a)}{H} &= \frac{\bar{H}}{\sqrt{1 + \bar{H}}},
\label{eq:meth03}
\end{align}
where $H$ is the scale height of the gas \citep{HP10}. The unitless number $\bar{H}$ is given by:\begin{align}
\bar{H} &= \left(\frac{1}{1 + \gamma_{turb}}\right)^{1/4}\sqrt{\frac{\alpha\Sigma_g}{\sqrt{2\pi}\rho_s a}},
\label{eq:meth04}
\end{align}
where $\rho_s$ is the bulk density of the dust, $\Sigma_g$ is the surface density of the gas, and $\gamma_{turb}$ is the exponent in the turbulent energy spectrum: $E(k)\propto k^{-\gamma_{turb}}$ where $k$ is the wavenumber. Generally $\gamma_{turb}$ has a value of between 5/3 and 3 \citep{Du95}, however our results are not sensitive to the choice of $\gamma_{turb}$ because of how weakly $\bar{H}$ depends on $\gamma_{turb}$ (see Equation \ref{eq:meth04}). We choose a value of 5/3 which is the classic Kolmogorov type of isotropic and incompressible turbulence. 

The above solution assumes that the grain stopping time $\tau_s \equiv \rho_s a / \rho c_s$ is vertically constant, and $\rho$ is the midplane gas density: $\rho \equiv \Sigma_g / \sqrt{2\pi} H$. This assumption is most relevant for large grains ($\Omega\tau_s \lesssim 1$) because they occupy the {\it strong settling limit} $H_d << H$ \citep{Fro09} and variations in $\tau_s$ can be neglected. For smaller grains ($\Omega\tau_s << 1$), the assumption breaks down because more grains are vertically extended and hence variations in $\tau_s$ become more important. The different vertical distributions do not result in major differences in the midplane ionization (see Appendix \ref{sec:append01}) because it depends on the total vertical optical depth of the disk which is not affected by settling. 

\cite{Fro09} have demonstrated the breakdown of the above assumption in global MHD simulations of stratified protoplanetary disks. In that work, they showed that the Gaussian shape for the vertical distribution of dust overestimates the amount of dust at $z > 2H$ when compared to their simulated distribution for all grain sizes. However the overestimation was worse for the smaller grains than for the large ones. \cite{Fro09} suggested that the deviation from Gaussian was due to a break down of the constant stopping time assumption as well as the stratified vertical structure of the gas in their simulations, which resulted in a turbulent diffusion coefficient that varied with height. 

The complexities that are afforded by a global MHD simulations are beyond the scope of this work, however we note that an overestimation of the density of dust at $z>2H$ implies an overestimation of the total opacity at those heights. This would results in a lower flux of ionization radiation at the midplane at large radii, especially early on in the disk lifetime when the dust is most extended. These discrepancies will be less important as the disk evolves and the gas density is reduced and more dust settles to the midplane.

\subsection{ Disk Chemistry and Ionization }

To compute the ionization structure we use a non-equilibrium photochemical code as described in \cite{Fog11} and CPA16. The code computes the astrochemistry for a disk irradiated by X-ray and UV radiation. Chemistry is primarily driven by ion-neutral reactions in the gas phase and is catalyzed by the presence of dust grains. Additionally, there are two grain surface reactions for the formation of water and molecular hydrogen. As well as catalyzing gas phase reactions, the grains impact the ionization structure by acting as a sink to electrons. Cosmic ray and X-ray ionization drives the chemistry by creating ions and free electrons. These free electrons also couple to the magnetic field which produces turbulence via the MRI.

We use a standard ionization rate of $10^{17} ~s^{-1}$ per H$_2$ for the cosmic rays and an assumed luminosity of $10^{30}$ erg/s for the X-rays. We model a range of X-ray wavelengths between 1 and 20 keV using a template spectrum for a T Tauri star. Some recent work (ex. \cite{Cle14,Cle15}) has suggested that disk observations are consistant with lower cosmic ray ionization rates, which could lead to a larger dead zone. While we do not explore this effect we do find that the dead zone location is sensitive to the size and density distribution of the dust, which suggests that the X-ray radiation field is having a significant impact on the gas ionization.

We compute an average dust grain size, weighted by the freezing efficiency parameter of the chemical code. The freezing efficiency parameter has the form: $fe(a) = (a/0.1\mu{\rm m})^{-3/2}$ and the average freezing efficiency is:\begin{align}
\bar{fe}(r) = \frac{\int^{a_{max}}_{a_{min}} n(a,r,z=0) fe(a) da}{\int^{a_{max}}_{a_{min}} n(a,r,z=0) da}
\label{eq:04b}
\end{align}
where $n(a,r,z=0)$ is the number density distribution at the midplane. Hence the size of the average grain is:\begin{align}
\bar{a}(r) = 0.1\mu{\rm m} \cdot \bar{fe}(r)^{-2/3}
\label{eq:04c}
\end{align}
In past works (eg. \cite{Fog11}) the freezing efficiency is assumed to be a global property of the disk, and variations in height and radii are ignored. Here we allow the freezing efficiency to vary as a function of radius while ignoring variations above the midplane. This underestimates the freezing efficiency of the grains at $z\gtrsim H$ which are smaller on average than at the midplane and could result in higher ionization at $z \gtrsim H$ because the efficiency of electron capture onto the grains is underestimated.

This average grain size and its freezing efficiency represent the average effect of the full size distribution in an attempt to minimize computational complexity. Every additional grain size added to the chemical model would require its own set of freezing, and grain surface reactions. For this reason, simply doubling the number of grains in the chemical model (two sizes instead of one) would nearly double the number of chemical reactions in the model. This larger set of reactions would require longer computation times to determine the final solution.

We have made no changes to the chemical code of \cite{Fog11}, and have tested our implementation with RADMC3D against some general astrochemical properties like condensation fronts (CPA16). Our method is to compute the dust surface density profiles for grain sizes between 0.1 $\mu$m to 200 cm in 100 bins that are spaced equally in logspace. We then interpolate the surface densities down to 20 bins, normalizing so that the total surface density is the same as before. On this smaller binned sample we compute the radiation field using RADMC3D. We use the results of the dust calculation to estimate the average grain size that is used by the chemical code.

The Ohmic Elsasser number is a unitless parameter that traces the level of ionization and is used to infer the location and extent of the dead zone \citep{Fro13}. It is defined as the ratio between the dissipation timescale and the growth timescale of the most unstable MRI mode. The critical value of this ratio which denotes the transition between turbulently dead and turbulently active regions is 1. This value is physically motivated as it represents an equilibrium between the growth and dissipation of turbulent energy \citep{Gammie96}. It has also been shown in numerical simulations that an Ohmic Elsasser number of 1 represents a transition point between strong MRI driven turbulence and either weak or decaying turbulent solutions \citep{T07}.

We connect the level of ionization from the results of the astrochemical code to the location of the dead zone using the Ohmic Elsasser number. It has the form of:\begin{align}
\Lambda_O \equiv \frac{\tau_{diss}}{\tau_{grow}} = \frac{v_{A,z}^2}{\eta \Omega}.
\label{eq:05}
\end{align}
In this expression the Alfv\'en speed in the z-direction is $v_{A,z} \equiv B_z/\sqrt{4\pi \rho}$, where $\rho$ is the gas density, $\Omega$ is the orbital frequency, $\eta$ is the Ohmic resistivity and $B_z$ is the local magnetic field in the z-direction \citep{Gres15,Crid16}. It has been shown \citep{BS11} that at equipartition in a protoplanetary disk the ratio of the gas pressure to the magnetic pressure (the `plasma beta', $\beta$) is related to the turbulent parameter $\alpha$ through: $\beta\sim 1/2\alpha$. Using this relation, we relate the local magnetic field to measurable fluid quantities\begin{align}
\beta \equiv \frac{P_{gas}}{P_{mag}} &\sim \frac{1}{2\alpha} \nonumber\\
\frac{2\cdot4\pi\rho c_s^2}{B^2} &\sim \frac{1}{2\alpha}\nonumber\\
B^2 &\sim 4\cdot4\pi\alpha\rho c_s^2
\label{eq:05x}
\end{align}
where we assume that the magnetic field energy is dominated by the z-component of the field. This has been observed to within a factor of order unity in simulations of MRI \citep{BS11}. This factor does not drastically change the location of the dead zone in our model. Finally the Ohmic Elsasser number is:\begin{align}
\Lambda_O = \frac{v^2_{A,z}}{\eta\Omega} = \frac{4c_s^2}{\eta\Omega}
\end{align}
The Ohmic resistivity is connected to the disk's ionization through \citep{KB04}: \begin{align}
\eta = \frac{234 T^{1/2}}{x_e} cm^2s^{-1},
\label{eq:06}
\end{align}
where $T$ is the temperature and $x_e$ is the electron fraction - the ratio of electrons to hydrogen atoms. We assume that a dead zone exists in regions of the disk where $\Lambda_O < \Lambda_{O,crit} \equiv 1$. In computing the Ohmic Elsasser number in Equation \ref{eq:05} we assume that the turbulent parameter is constant throughout the disk. Reducing $\alpha$ at radii where $\Lambda_O < \Lambda_{0,crit}$ does not change our inference of where the dead zone edge is located.

In discussing a varying turbulent $\alpha$ it should be pointed out that there is a difference between the {\it turbulent} $\alpha$ and the {\it effective viscosity} (EV) $\alpha$ , which we assume is constant. The turbulent $\alpha$ describes the rate of angular momentum transfer due to turbulence. While the EV $\alpha$ describes the angular momentum transport through the disk, which can be caused by turbulence, disk winds, and spiral density waves.

A constant EV $\alpha$ is a current limitation of our disk model, and hence we assume that $\alpha_{EV} \equiv \alpha_{turb} + \alpha_{wind} = {\rm constant}$. This assumption corresponds to a constant (in radius) mass accretion rate throughout the disk. 
\section{ Results }\label{sec:results}

\begin{figure}
\centering
\subfigure[t = 0.1 Myr]{
	\label{fig:res01a}
	\includegraphics[width=0.5\textwidth]{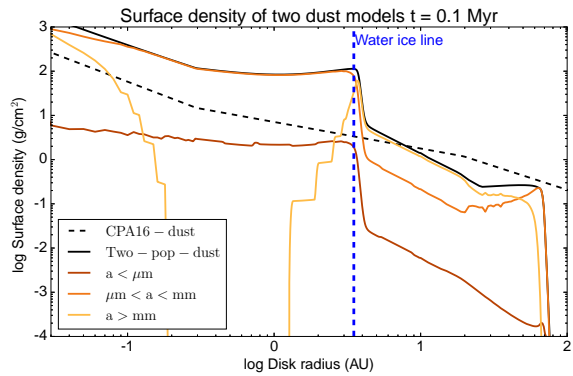}
}
\subfigure[t = 1.3 Myr]{
	\label{fig:res01b}
	\includegraphics[width=0.5\textwidth]{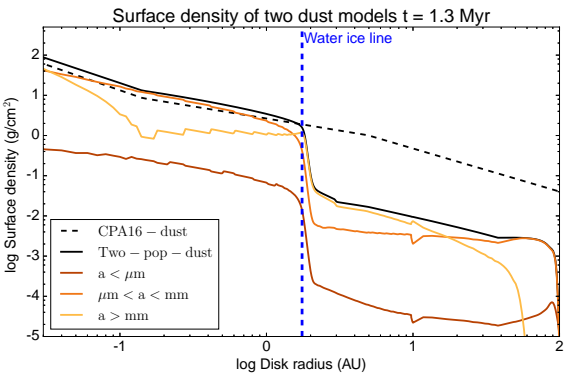}
}
\subfigure[t = 3.7 Myr]{
	\label{fig:res01c}
	\includegraphics[width=0.5\textwidth]{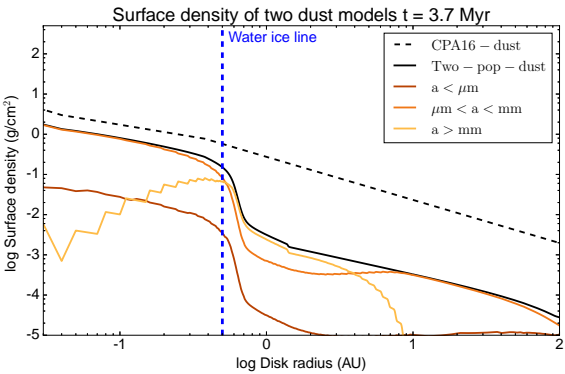}
}
\label{fig:res01}
\caption{ Total surface density (solid and dashed black lines) for the two dust models that were investigated. The coloured lines represent the total surface density of binned dust grain sizes that has been estimated by the Two-pop-dust model. The blue dashed line denotes the location of the water ice line. The largest grains show a depletion within the ice line where fragmentation becomes more efficient. }
\end{figure}

\begin{figure}
\centering
\subfigure[t = 0.1 Myr]{
	\label{fig:res02a}
	\includegraphics[width=0.5\textwidth]{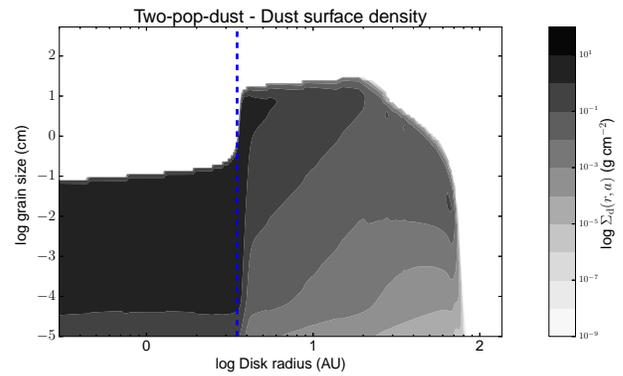}
}
\subfigure[t = 1.3 Myr]{
	\label{fig:res02b}
	\includegraphics[width=0.5\textwidth]{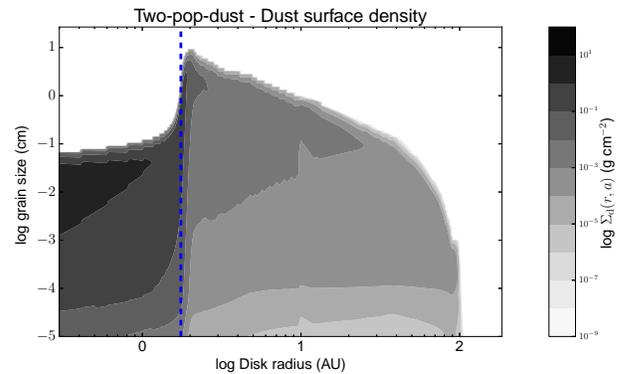}
}
\subfigure[t = 3.7 Myr]{
	\label{fig:res02c}
	\includegraphics[width=0.5\textwidth]{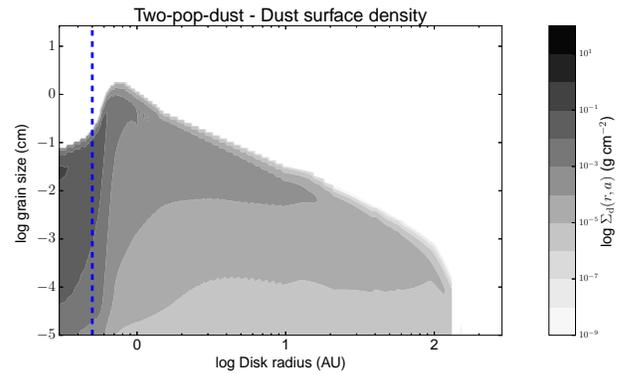}
}
\label{fig:res02}
\caption{ 2D representation of the estimated dust surface density for the different grain sizes in the Two-pop-dust model as a function of radius. We can clearly see a drop in the maximum grain size of over two orders of magnitude at the ice line (denoted by the blue dashed line).}
\end{figure}

\begin{table}
\begin{center}
\caption{ CPA16 disk model parameters }
\begin{tabular}{c c l}\hline
Parameter & Value & Notes\\\hline
$M_{disk}(t=0)$ & $0.1$ M$_\odot$ & Initial disk mass \\\hline
$s_{disk}(t=0)$ & $66$ AU & Initial disk outer radius \\\hline
$M_{star}$ & $1.0$ M$_\odot$ & Stellar mass \\\hline
$R_{star}$ & $3.0$ R$_\odot$ & Stellar radius \\\hline
$T_{star}$ & $4200$ K & Stellar effective temperature \\\hline
$\alpha$ & $0.001$ & Turbulent parameter \\\hline
$t_{life}$ & $4.10$ Myr & Disk lifetime \\\hline
$L_{xray}$ & $10^{30}$ erg/s & Total X-ray luminosity\\\hline
\end{tabular}\\
\label{tab:result01}
\end{center}
\end{table}

To demonstrate the impact of the dust distribution on the ionization structure we compute the radiation field, astrochemistry, and ionization in the CPA16-dust and Two-pop-dust models at 0.1, 1.3 and 3.7 Myr for the fiducial gas disk model from \cite{Crid16} (see Table \ref{tab:result01} for disk model parameters). These times are meant to represent the disk at its early, middle and late ages.

\subsection{ Dust Surface Density Radial Distribution }

In Figures \ref{fig:res01a}, \ref{fig:res01b} and \ref{fig:res01c}  we show the radial dependence of the total surface density of the dust in both models. We find the same qualitative behaviour as previous works where the surface density is enhanced at the ice line (denoted by the blue dotted line) by more than an order of magnitude. This enhancement is caused by a lower maximum grain size within the ice line because the fragmentation threshold speeds have dropped by an order of magnitude. The enhancement of the dust surface density and the retention of the dust is important for planet formation as it results in higher accretion rates during the early stages of solid accretion which will impact the final mass and evolution of forming planets. We will explore this particular facet of this problem in a later paper.

Figures \ref{fig:res01a}, \ref{fig:res01b} and \ref{fig:res01c} also show the radial distribution of different grain sizes that are inferred by the Two-pop-model by reconstructing the size distribution based on the results of \cite{B15}. In the figure, the grain sizes have been binned into sub-micron, micron to sub-millimeter and greater than millimeter. We find that the large grains are depleted within the ice line due to fragmentation and radial drift, while they tend to dominate the surface density at large radii and early times. As the disk ages these large grains radially drift inwards, depleting the solid density at the most distant radii. This depletion moves large grains to smaller radii, within the ice line, where they replenish the surface density of the large grains. Within the ice line the solid mass is dominated by micron to millimeter sized grains at all times. This size range also dominates the dust mass at large radii and late times.

In Figures \ref{fig:res02a}, \ref{fig:res02b} and \ref{fig:res02c} we show the dust surface density (gray scale) as a function of grain size and disk radius. On the gray scale, white denotes the region of the plot where a grain size is not populated by any surface density. We see that the maximum grain size that is populated by surface density drops by up to two orders of magnitude across the water ice line (blue dashed line). The radial dependence of the maximum grain is set by Equations \ref{eq:02} and \ref{eq:03} and depends on which process limits the maximum grain. This drop in grain size is the important aspect that leads to the longer retention of dust in protoplanetary disks. \cite{B12} showed that the radial drift timescale $\tau_{drift} = r / u_D \sim 1 / a$, which implies that a drop in the maximum grain size of two orders of magnitude leads to an increase in the radial drift timescale of two orders of magnitude. This increase in the radial drift timescale is a feature of the dust physics that can 
lead to the longer retention of the dust in protoplanetary disks, which impacts the amount of material available 
for planetary accretion.

\begin{figure}\label{fig:res07}
\centering
\subfigure[t = 0.1 Myr]{
	\label{fig:res07a}
	\includegraphics[width=0.5\textwidth]{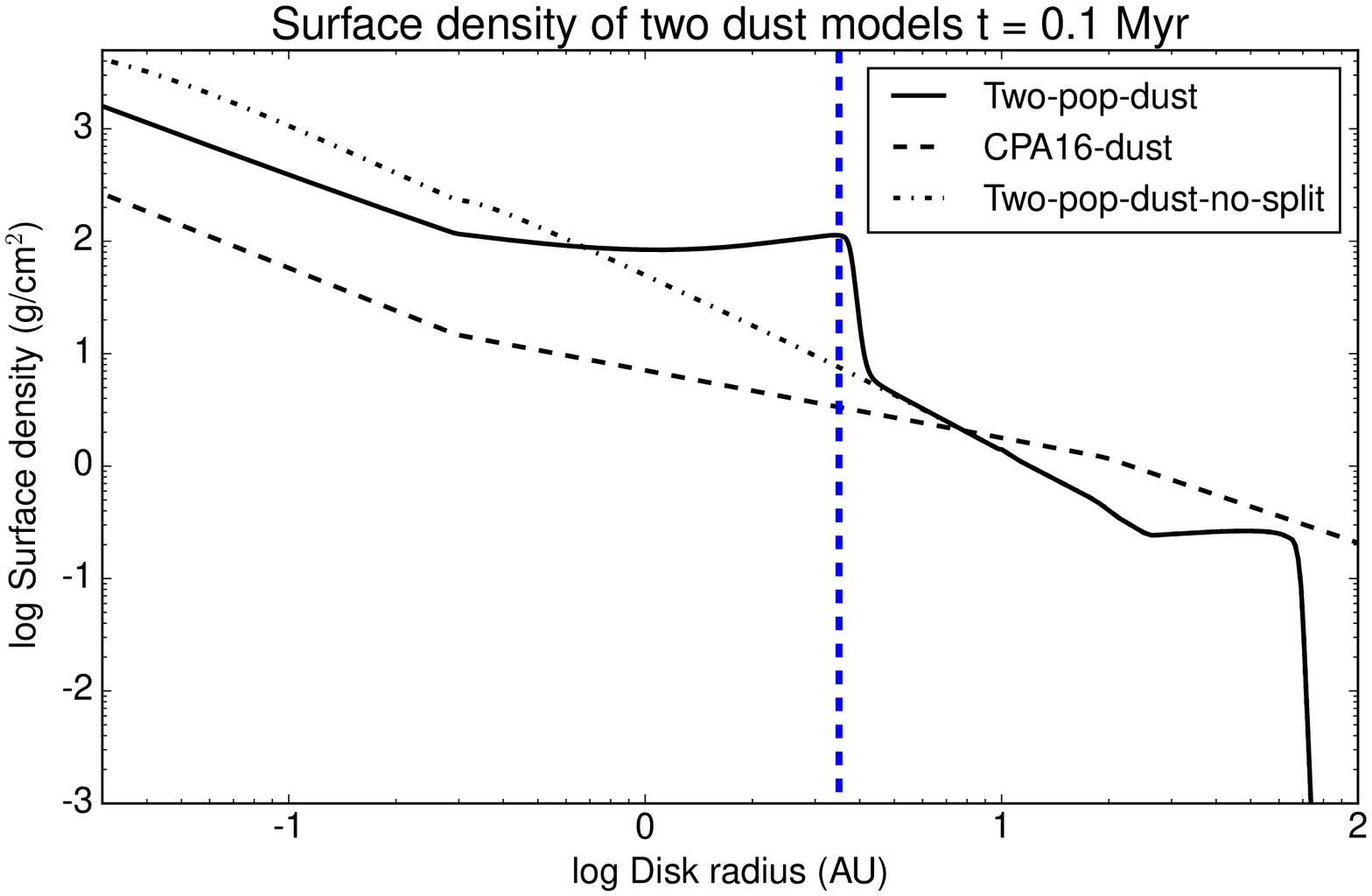}
}
\subfigure[t = 1.3 Myr]{
	\label{fig:res07b}
	\includegraphics[width=0.5\textwidth]{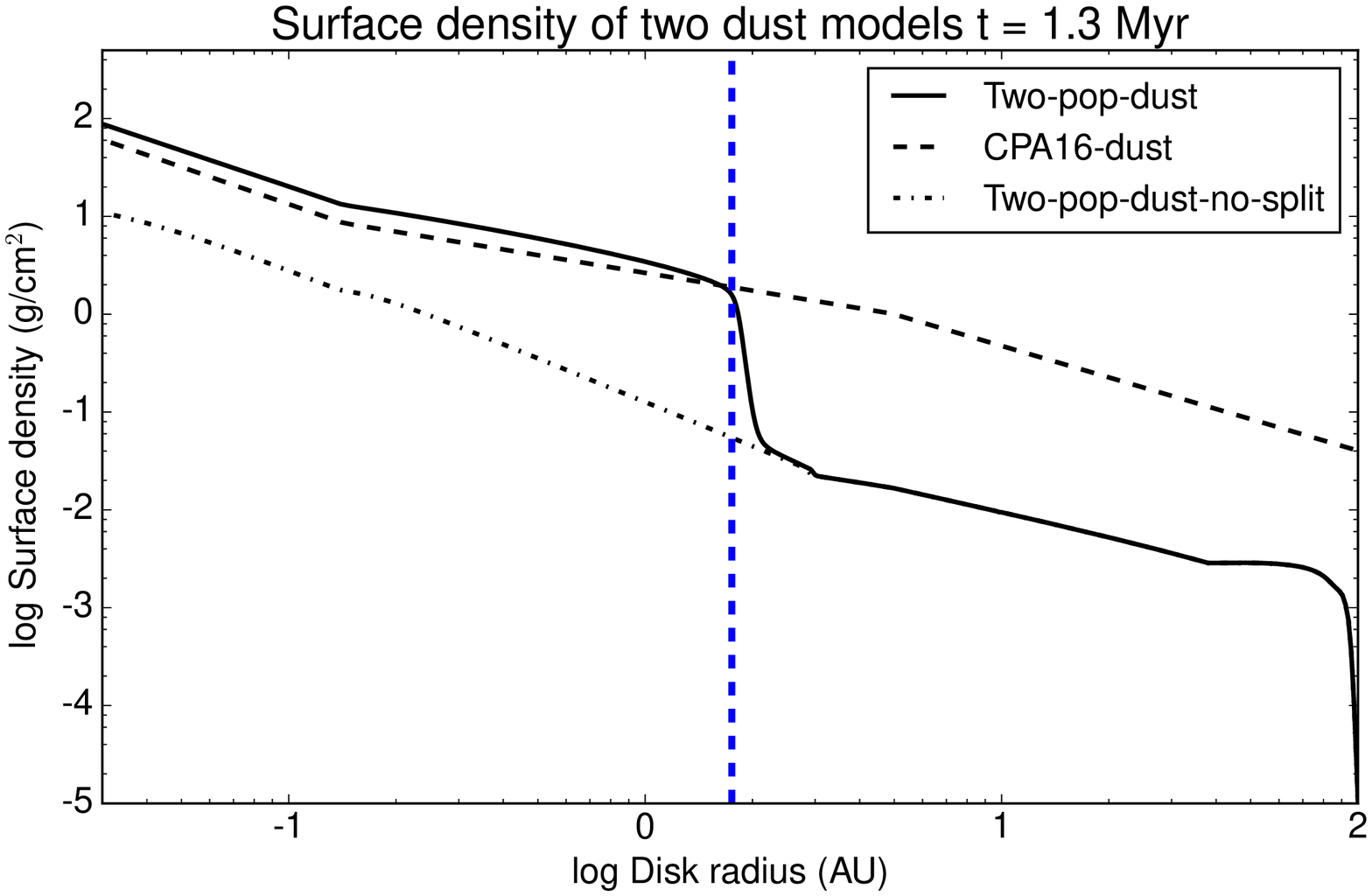}
}
\subfigure[t = 3.7 Myr]{
	\label{fig:res07c}
	\includegraphics[width=0.5\textwidth]{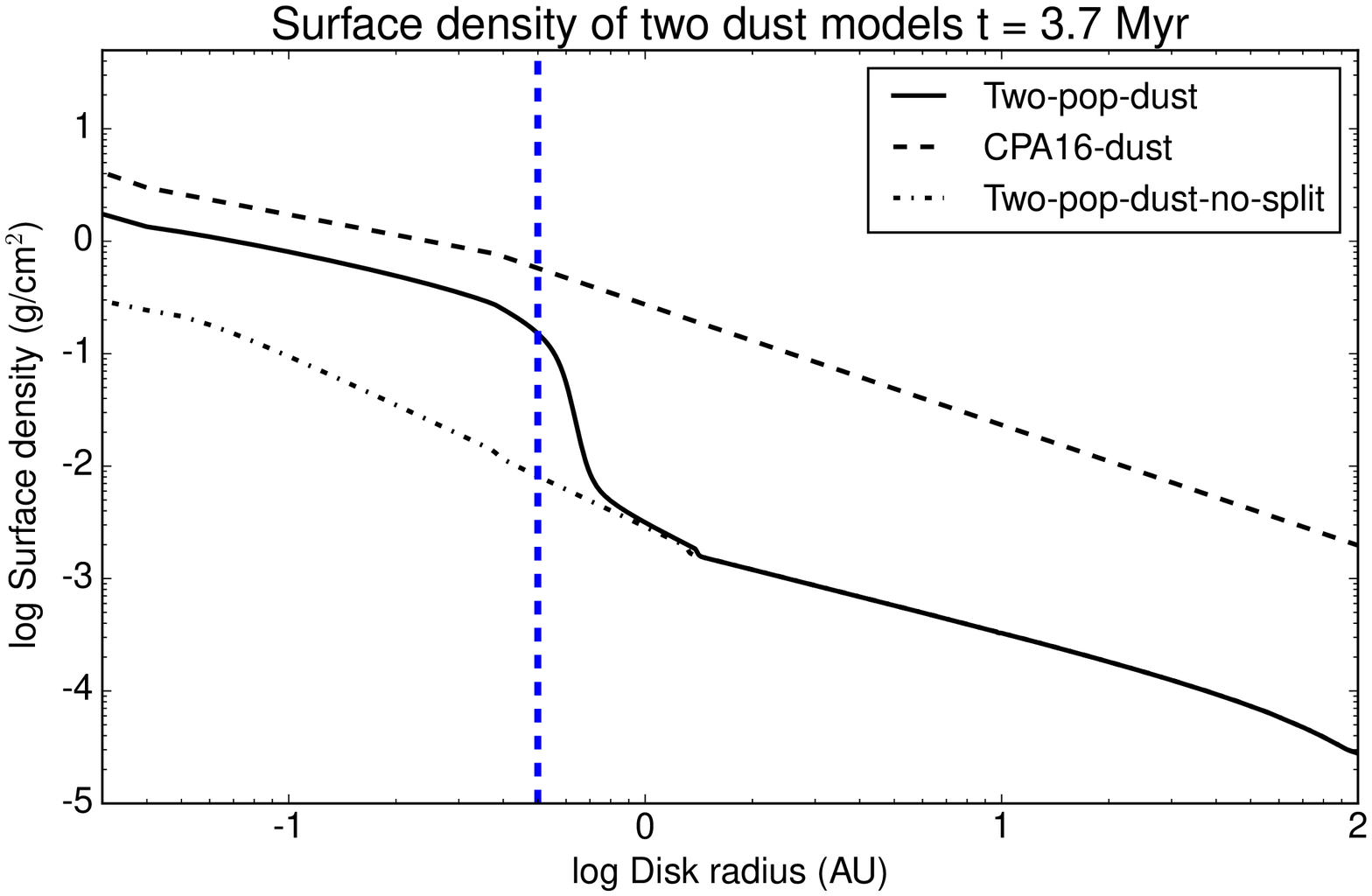}
}
\caption{ Radial dependence of total dust surface density in the CPA16-dust (dashed), Two-pop-dust (solid), and Two-pop-dust-no-split (dotted) models at 0.1, 1.3, and 3.7 Myr. The Two-pop-dust-no-split model is the same as the Two-pop-dust model except that we have not changed the fragmentation threshold speed at the ice line. Instead, it is kept constant at 10 m/s.  }
\end{figure}

\begin{figure}
\centering
\includegraphics[width=0.5\textwidth]{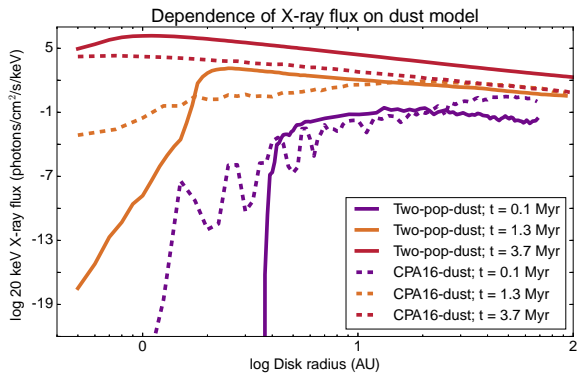}
\caption{ Radial distribution of the 20 keV X-ray photons at the midplane of the disk for the two dust models at 0.1, 1.3, and 3.7 Myr. The at earlier times Two-pop-dust model has higher surface density at lower radii and generally truncates the radiation field at larger radii than in the CPA16-dust model. Later on the dust in the Two-pop-dust model has cleared out due to radial drift, which results in a higher flux of X-rays. }
\label{fig:res03}
\end{figure}

\begin{figure}
\centering
\subfigure[ In the CPA16-dust model the disk is well shielded for its entire history. The dead zone begins large and does not shrink far as the disk ages.  ]{
\includegraphics[width=0.5\textwidth]{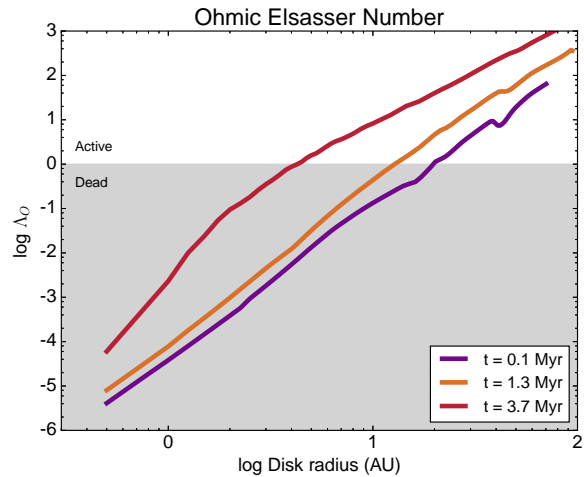}
\label{fig:res04a}
}
\subfigure[ In the Two-pop-dust model the dust is evacuated by radial drift much faster than by viscous stresses. This results in a higher ionization on the midplane and a smaller dead zone at later times. ]{
\includegraphics[width=0.5\textwidth]{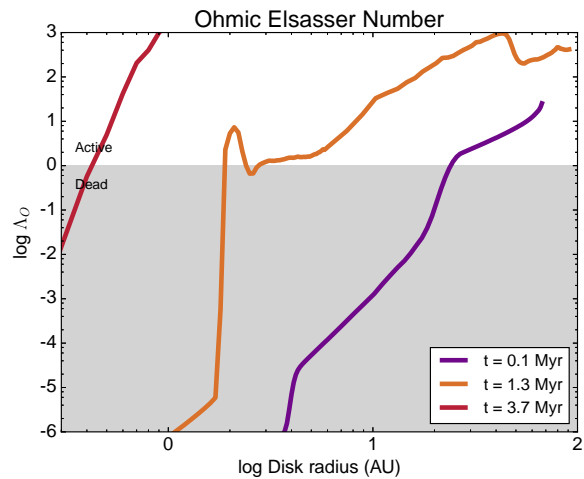}
\label{fig:res04b}
}
\caption{ Evolution of the Ohmic Elsasser number for the two dust models. The gray region shows the range of Ohmic Elsasser numbers where we expect the viscosity parameter due to turbulence is lower by two order of magnitudes than in the case of the white region. }
\label{fig:res04}
\end{figure}

\subsection{ Dust Retention }

To demonstrate that the ice line is indeed causing a higher retention of the dust, we ran the Two-pop-dust model without changing the threshold speed at the ice line (Two-pop-dust-no-split model). 

We show the results of this test in Figures \ref{fig:res07a}-\ref{fig:res07c}. In these figures we plot the surface density of dust as a function of radius for the CPA16-dust (dashed), Two-pop-dust (solid), and Two-pop-dust-no-split (dotted) models. We find that the dust dissipates faster in the Two-pop-dust-no-split model than in CPA16-dust. This is due to the radial drift dominating the dust evolution, causing rapid inward migration of the largest dust grains. This impacts the viability of forming planets through core accretion, as the accretion timescale depends on the amount of dust available to accrete. The maximum grain size dictates how efficiently the dust is cleared out because the larger grains are more susceptible to radial drift. This connection implies a link between the treatment of the fragmentation threshold speed and the viability of forming planets through core accretion. If the dust grains are too large and disappear too quickly there will not be 
enough material to build the solid cores that lead to Jupiter-mass planets.

We do find that the dust in the outer disk is cleared out very rapidly. Decreasing from $\sim 0.001 M_\odot$ at 0.1 Myr down to $\sim 10^{-7} M_\odot$ by 3.7 Myr. Such a reduction is generally inconsistant with observations which show that disks can remain dust-rich for millions of years \citep{Nat07,Ricc10}. Some methods such as gas pressure maxima (ex. \cite{Pin12}) have been suggested as dust traps at large radii. In principle, the sources of these pressure maxima could be similar to the sources of planet traps (eg. dead zone edge, ice lines of volatile species other than H$_2$O), however their effects are not included in this work.

\subsection{ Midplane X-ray Flux }

In Figure \ref{fig:res03} we show the radiation field along the midplane for the two dust models at 0.1, 1.3, and 3.7 Myr. In the Two-pop-dust model the dust surface density is higher at 0.1 and 1.3 Myr, truncating the radiation field along the midplane at larger radii than in the CPA16-dust model. Early in the disk life the flux of radiation at the midplane of the Two-pop-dust model is more strongly truncated than in the CPA16-dust model. This truncation is due to the dust enhancement that occurs within the ice line and leads to low levels of ionization in that region of the disk. At 1.3 Myr the dust in the Two-pop-dust model is cleared outside the ice line by radial drift. This results in a radiative flux that is three orders of magnitude higher near the ice line than in the CPA16-dust model. Finally at the latest time, the dust surface density has been cleared out by radial drift and the resulting flux of X-rays is higher everywhere in the disk.

\begin{figure}
\centering
\subfigure[t = 0.1 Myr - CPA16-dust]{
	\label{fig:res05a}
	\includegraphics[width=0.5\textwidth]{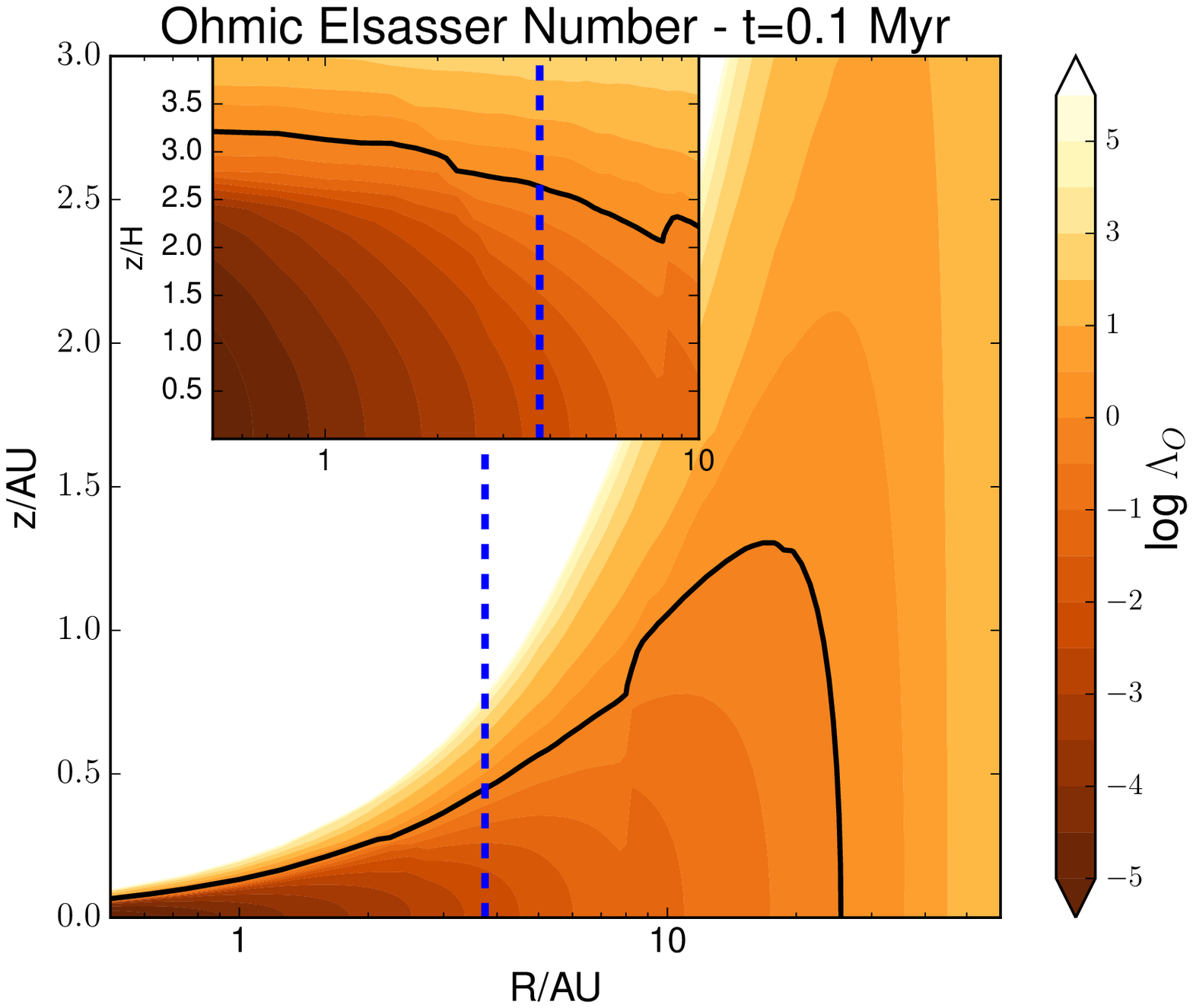}
}
\subfigure[t = 1.3 Myr]{
	\label{fig:res05b}
	\includegraphics[width=0.5\textwidth]{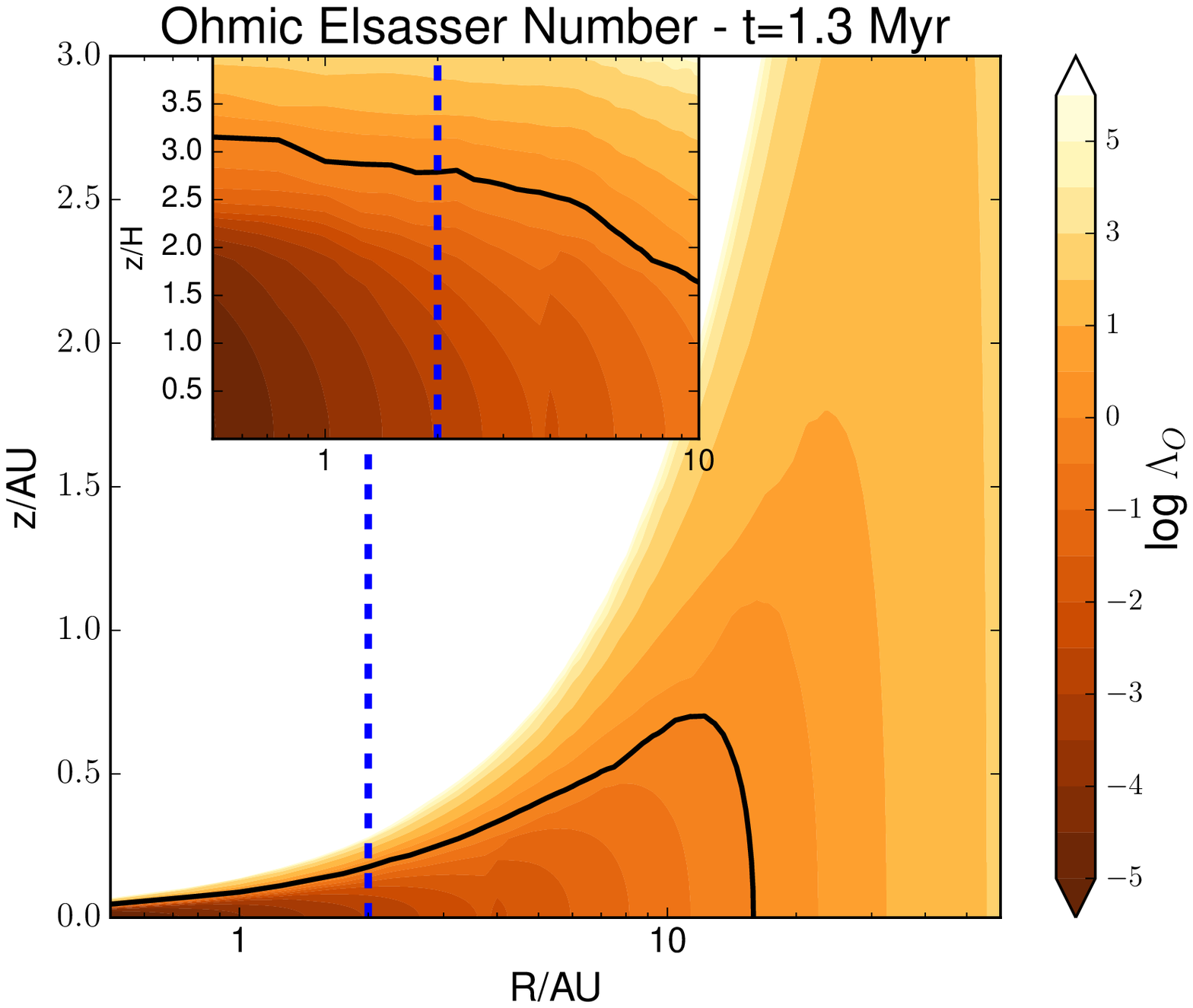}
}
\subfigure[t = 3.7 Myr]{
	\label{fig:res05c}
	\includegraphics[width=0.5\textwidth]{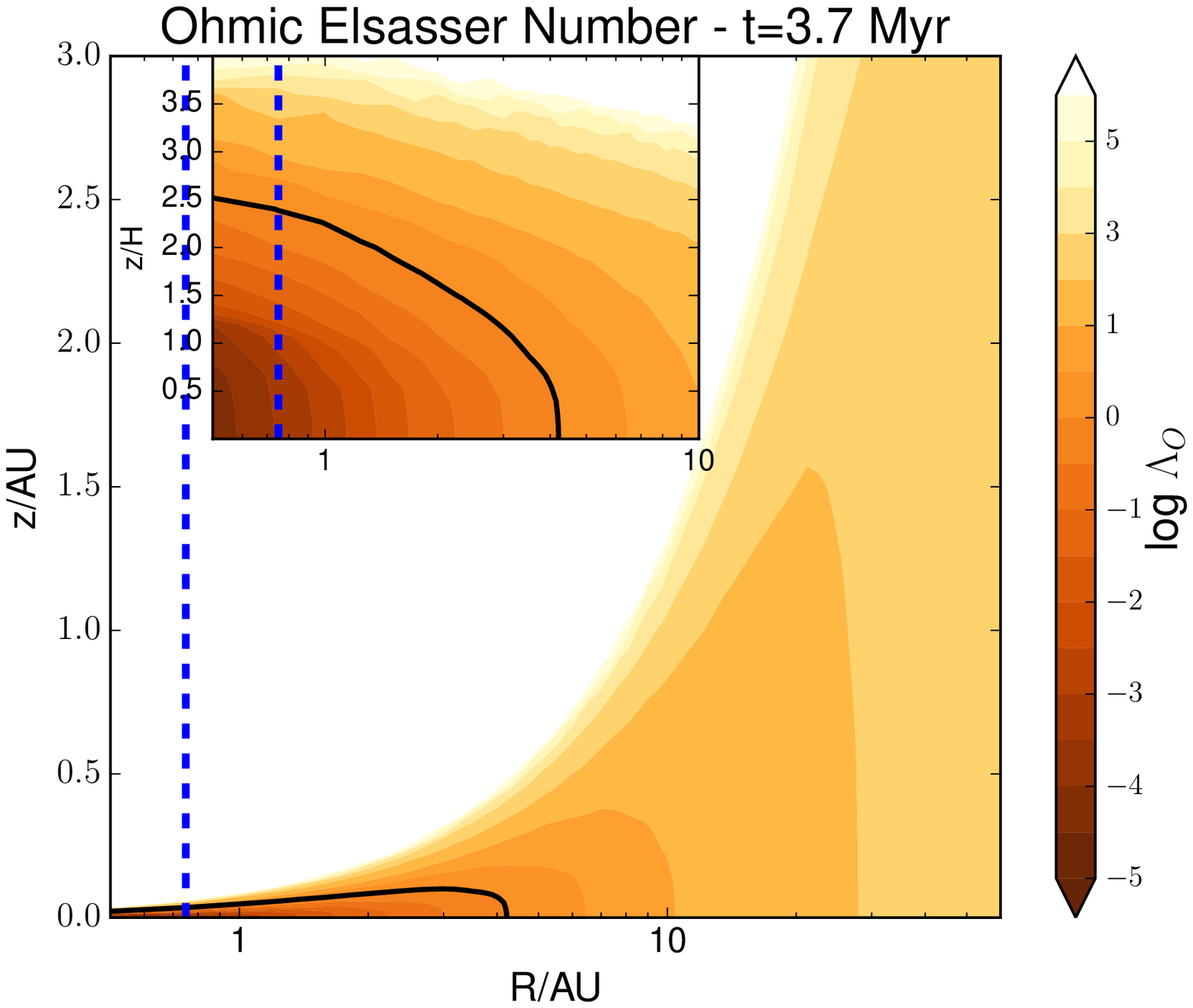}
}
\label{fig:res05}
\caption{ Radial and height distribution of Ohmic Elsasser number for our CPA16-dust model. The blue dashed line shows the approximate location of the water ice line. The solid black line shows $\Lambda_O = \Lambda_{O,crit}\equiv 1$. The inset scales the height by the gas scale height. }
\end{figure}

\begin{figure}
\centering
\subfigure[t = 0.1 Myr - Two-pop-dust]{
	\label{fig:res06a}
	\includegraphics[width=0.5\textwidth]{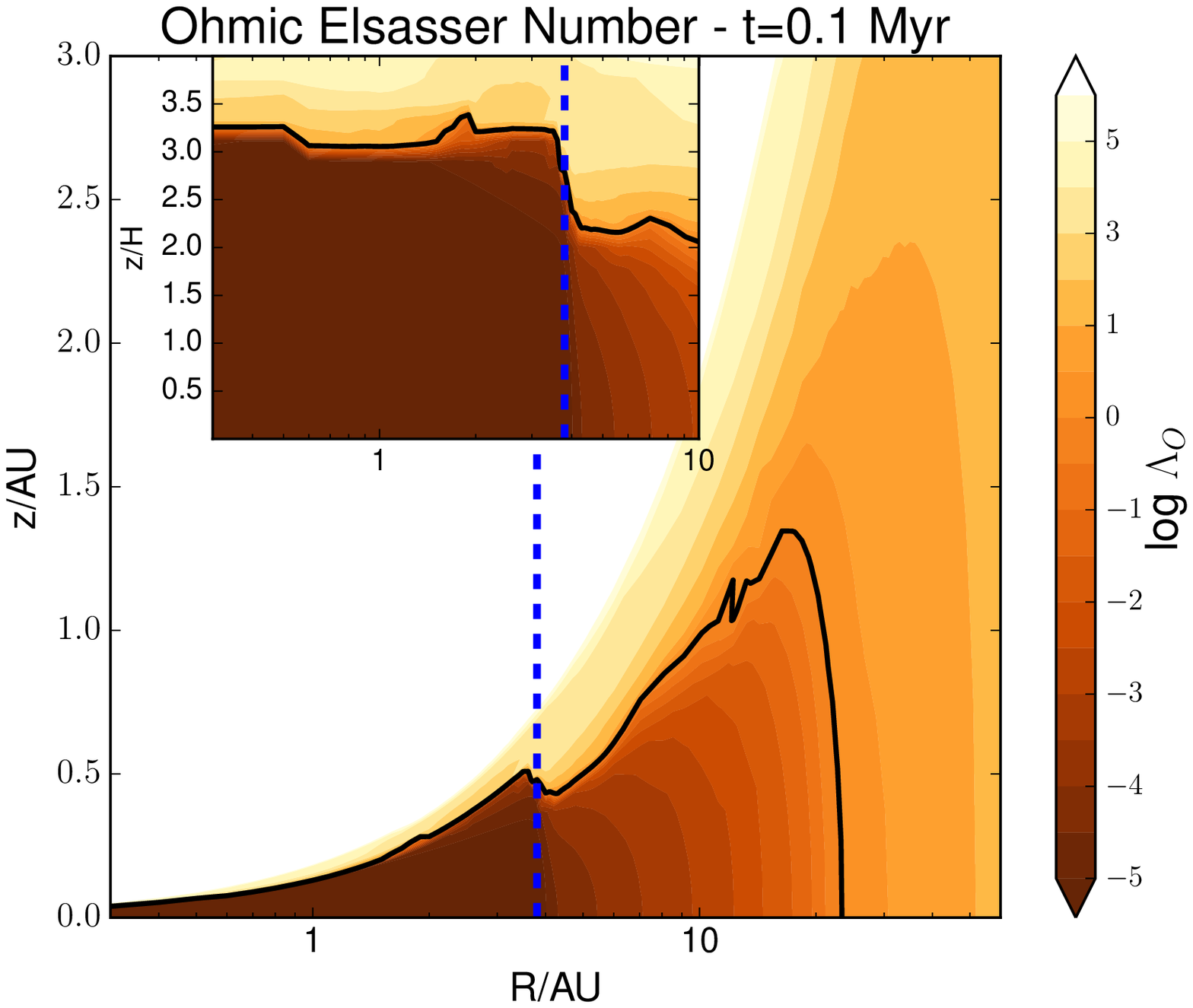}
}
\subfigure[t = 1.3 Myr]{
	\label{fig:res06b}
	\includegraphics[width=0.5\textwidth]{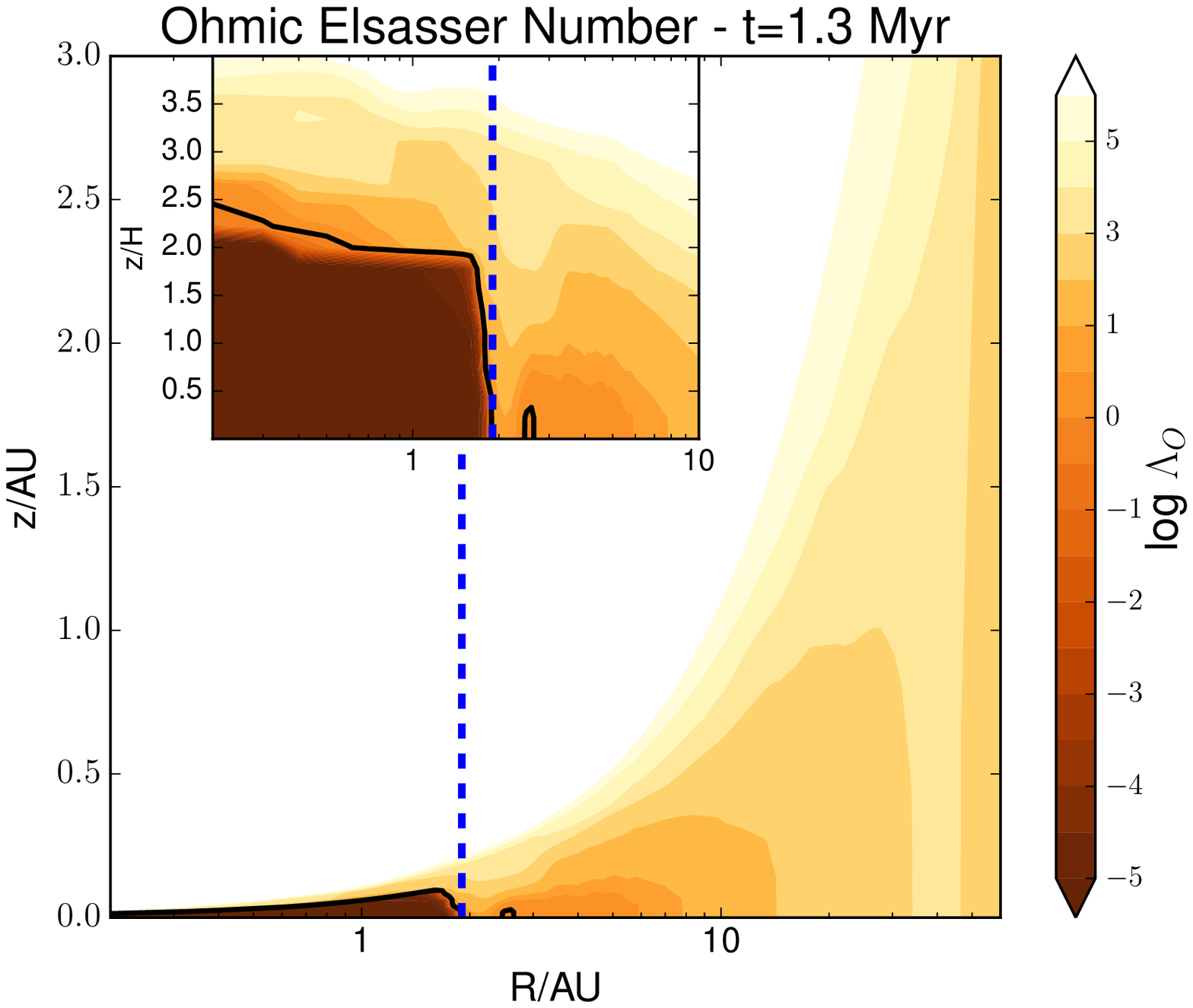}
}
\subfigure[t = 3.7 Myr]{
	\label{fig:res06c}
	\includegraphics[width=0.5\textwidth]{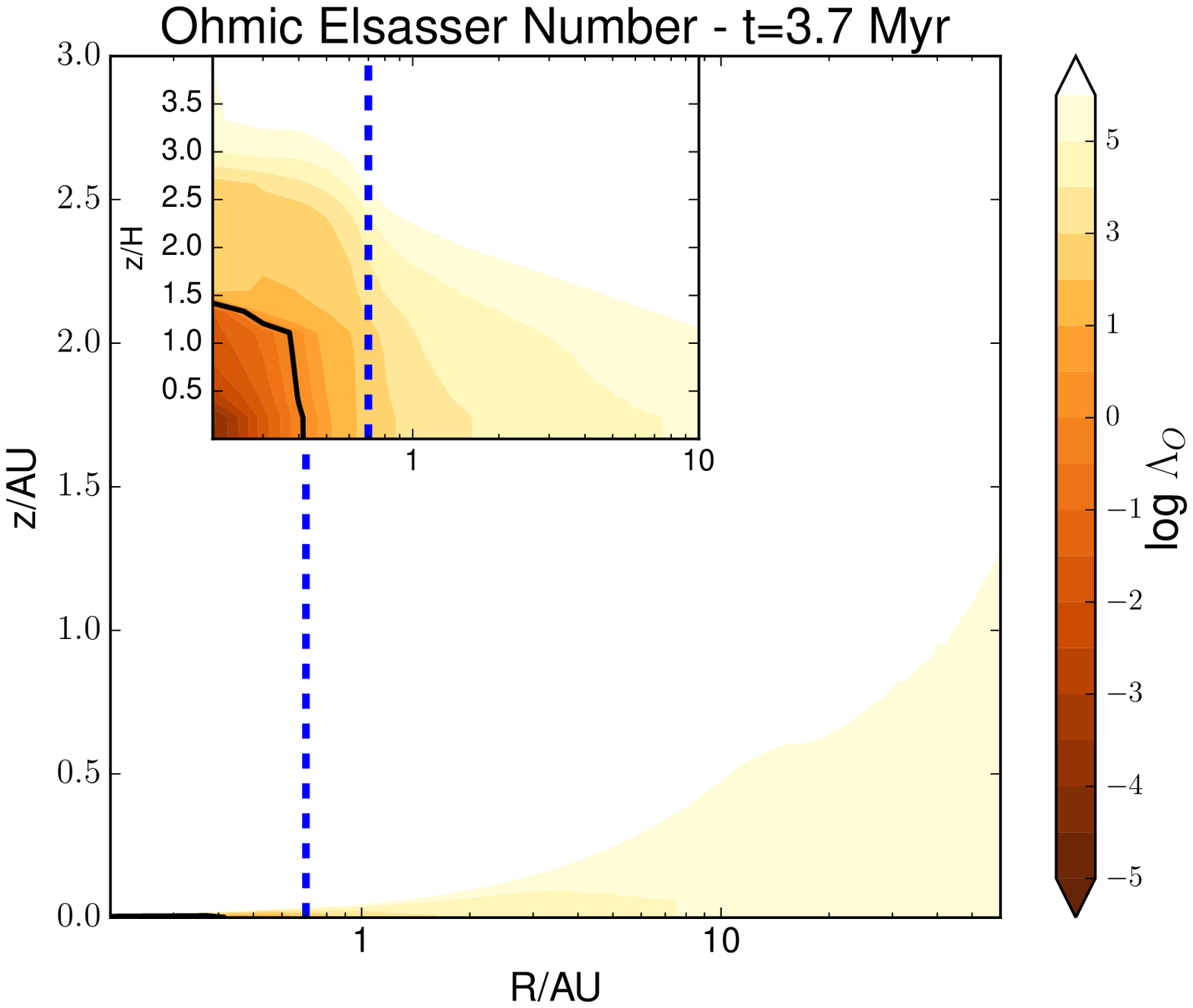}
}
\label{fig:res06}
\caption{ Same as in Figures \ref{fig:res05a}-\ref{fig:res05c} for our Two-pop-dust model. The location of the ice line is approximately the same, accept within 1 AU because the resolution is higher in this model. The inset scales the height in units of the gas scale height.}
\end{figure}

\subsection{ Dead Zone Radial Evolution }

In Figures \ref{fig:res04a} and \ref{fig:res04b} we show the evolving radial profile of the Ohmic Elsasser number, computed from the results of our astrochemical code, along the disk midplane for the two dust models. The biggest difference between the two dust models is how quickly the outer edge of the dead zone evolves inward. In the CPA16-dust model the dead zone edge starts at $\sim 21$ AU and evolves to $\sim 4.5$ AU by 3.7 Myr. Meanwhile in the Two-pop-dust model the dead zone edge starts at $\sim 24$ AU and evolves to $\sim 0.4$ AU by 3.7 Myr.  This order of magnitude increase in the rate of inward evolution for the dead zone edge is caused by the more rapid evolution of the dust in the new model. In the CPA16-dust model the dust surface density evolves only through the accretion of material onto the star through viscous stresses. This is generally a slow process, with evolution timescales on the order of a few million years. Conversely, in the Two-pop-dust model the dust evolution is dominated by a 
much faster radial drift. The dust starts at a higher surface density than in the CPA16-dust model, so the Ohmic Elsasser number (and hence the ionization) does start lower at the smaller radii in the Two-pop-dust model at $t = 0.1$ Myr. The lower ionization is quickly erased as the dust is cleared out through radial drift, and hence near the end of the disk lifetime the height of the dead zone is an order of magnitude lower than in the CPA16-dust model. 

Additionally, we find that within the ice line ($R\sim 2$ AU at 1.3 Myr) the ionization significantly drops off. As we have seen, inside the ice line the average grain is smaller than outside the ice line. These smaller grains have more surface area per mass and hence a higher electron capture rate. This directly impacts the ionization because free electrons tend to be much rarer.

\subsection{ 2D Structure of the Dead Zone }

In Figures \ref{fig:res05a}-\ref{fig:res05c} we show the radial and height dependence of the Ohmic Elsasser number in the CPA16-dust model at three times. The black contour shows edge of the dead zone where $\Lambda_O = 1$ and the blue dotted line shows the location of the water ice line. In the inset, we have focused on the the inner 10 AU of the disk and changed the units of the y-axis into gas scale heights. The purpose of this change in units is to illustrate the effect of dust settling, which is absent in the CPA16-dust model but present in the Two-pop-dust model. In the CPA16-dust model the edge of the dead zone moves to lower radii, while the height of the dead zone is nearly constant at lower radii.

Figures \ref{fig:res06a}-\ref{fig:res06c} are the same as in Figures \ref{fig:res05a}-\ref{fig:res05c} but plot the results for the Two-pop-dust model. In this case the dead zone not only evolves radially faster than in the CPA16-dust model, but it also tends to shrink towards the midplane faster. From the earlier to later snapshots of the disk the height of the dead zone reduces from 3 scale heights down to 1 scale height within the ice line. This shrinking contrasts the CPA16-dust model where the height of the dead zone does not change significantly. These results are connected to the settling of the largest dust grains which reduces the opacity of the disk higher in its atmosphere. This dependence on the dust settling also means that the ice line impacts the vertical structure of the disk ionization. 

At the ice line (denoted by the blue dashed lines in the figures) we find that the height of the dead zone drops by at least a gas scale height when the disk surface density is high (Figure \ref{fig:res06a}) and by more when the dust density has dropped (ie. Figure \ref{fig:res06b}). At smaller radii than the orbital radius of the ice line the dust is, on average, smaller than the dust at larger radii. These smaller grains are less settled, leaving more dust mass at $z > 1-3 H$. This results in a more opaque disk at these heights, and lower ionization higher up in the disk.

We emphasize that the black contours outlining the height of the dead zone assumes that the Ohmic resistivity is the only restricting factor to the growth of the MRI modes. This non-ideal effect should dominate the highest densities of the disk and hence will most accurately model the location of the dead zone close to the midplane. In less dense regions of the disk, the MRI is sensitive to other non-ideal MHD effects. These non-ideal effects are ambipolar diffusion, which dominates in the most diffuse parts of the disk, and the Hall effect which should dominate somewhere between the other two \citep{BS13}. Because of our interest in planet formation, we focus on Ohmic resistivity as we expect the majority of planet formation to occur on along the midplane of the disk. The reader is cautioned that a more realistic dead zone likely extends to a higher disk height than is shown here when ambipolar diffusion and the Hall effect are included.

\section{ Discussion and Conclusions }\label{sec:dis}

\subsection{ Implications for Planet Formation }\label{sec:plntForm}

We have demonstrated that the radial evolution of the dead zone is sped up by an order of magnitude when more complicated dust physics is included. The evolution of the dead zone impacts the location of the forming planet that is trapped at its edge. A planet trapped at the dead zone in the Two-pop-dust model will form in regions of the disk that are higher in density than the planet that formed in the CPA16-dust model. In CPA16-dust model, the dead zone planet failed to form a Jovian mass planet because its accretion rate was too low over the entire 4.1 Myr of evolution. We expect that the planet forming in the dead zone trap in the Two-pop-dust model will achieve a higher mass because it will sample regions of the disk with a higher density of gas and dust, however we leave this calculation to an upcoming paper.

A second effect of the new dead zone evolution is on the saturation of the co-rotation torque, and the trapping power of the heat transition planet trap. This trap is located at the transition point where the primary heating mechanism changes from viscous stresses to direct irradiation (CPA16). This trap starts at 19.5 AU in the fiducial model of CPA16 and remains within the dead zone for the entire lifetime of the disk. Because it is within the dead zone, the planet forming in the heat transition trap saturates after 1.1 Myr and ends up as a Hot Jupiter, very close to its host star. 

\begin{figure}
\centering
\subfigure[ CPA16-dust trap evolution ]{
	\label{fig:dis01a}
	\includegraphics[width=0.5\textwidth]{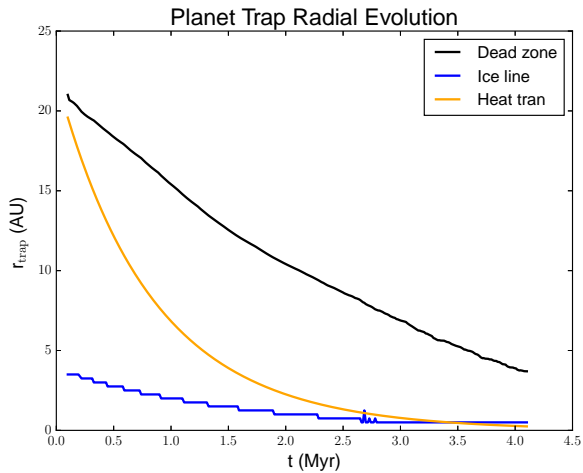}
}
\subfigure[ Two-pop-dust trap evolution ]{ 
	\label{fig:dis01b}
	\includegraphics[width=0.5\textwidth]{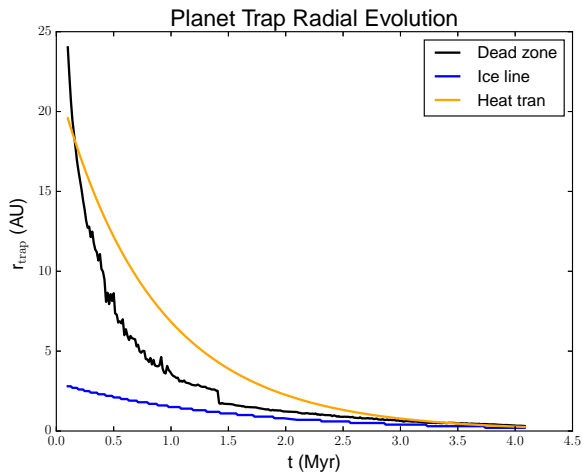}
}
\label{fig:dis01}
\caption{ Temporal evolution of the orbital radius of the three planet trap in our planet formation model. The heat transition trap evolves the same in both models because it is tied to the disk model. The ice line is nearly the same, while the dead zone evolution changes drastically. The step function evolution of the ice line is dependent on the resolution of the chemistry calculation, which is higher in the case of the Two-pop-dust model.}
\end{figure}

In Figures \ref{fig:dis01a} and \ref{fig:dis01b} we show the radial evolution of the three planet traps from our planet formation model. The location of the heat transition trap is defined by our gas disk model, and is the same in both dust models. The water ice line also shows little difference between the two models, which confirms the validity of using the results of CPA16 to parameterize the location of the ice line for the Two-pop-dust model. The evolution of the dead zone location shows the biggest change, confirming the results of the previous section. Because of the rapidly evolving dust, the dead zone edge shrinks toward the location of the ice line within 1.5 Myr. The rapid decrease at $t\sim 1.5$ Myr is linked to the radial structure of the midplane Ohmic Elsasser number at $r\sim 2$ AU at 1.3 Myr in Figure \ref{fig:res04b}. This feature is characterized by a peak near the location of the ice line, followed by a valley at larger radii, then a steadily increasing Ohmic Elsasser number with 
radius. As the disk ages, and the ionization increases, the location of the dead zone trap `hops' from the outside edge of the valley to its inner edge.

The heat transition trap and dead zone cross very early in the simulation, hence the saturation of the co-rotation torque will not occur at the same time as it did in the results of CPA16. This will keep the planet forming in the heat transition trap farther out in the disk, which slows its growth due to the lower density of material and changes the chemical content of its accreted atmosphere - if it has enough time to accrete an atmosphere.

This known evolution of the dead zone edge location is also important to the evolution of the dust grains because of the dependence of properties like fragmentation and settling on the amount of turbulence in the gas. While not explicitly stated in the methods section, we assumed a constant turbulent parameter throughout the disk that modelled a turbulently active fluid ($\alpha = 10^{-3}$). In the Appendix we fit the location of the dead zone as a function of time from Figure \ref{fig:dis01b} to model the temporal evolution of the radial dependence of the turbulent parameter. We find that the evolving turbulent parameter does not drastically change the resulting location of the dead zone at 0.1 Myr (see Appendix).

\ignore{At 1.3 Myr, the heat transition trap is at 4.8 AU, which is at a larger radius than the location of the dead zone in the Two-pop-dust model at the same time. Because the dead zone is at smaller radii than the heat transition trap, they must have crossed at some point between 0.1 and 1.3 Myr. The timing of this crossing is very important for the trapping of the planet forming in the heat transition trap. If the dead zone edge crosses within the radial location of the heat transition trap before the co-rotation torque saturates then the saturation will not occur and the planet will remain trapped for the remainder of the disk lifetime. The timing, and its effect will be quantified in an upcoming paper.}

\subsection{ Implications for structure seen in ALMA observations of disks }

The increase in water vapour and reduction of maximum grain size at radii inside the location of the ice line should have an observable effect on the continuum emission as observed by ALMA \citep{Ban15}. It has been suggested (ie. in \cite{And16}) that continuum emission gaps in the observations of the protoplanetary disk with ALMA could be explained by the presence of an ice line. In this scenario the increase in water vapour at lower radii than the location of the ice line modifies the opacity which affects the continuum emission. 

The effects of the ice line have recently been directly observed with ALMA in the V883 Ori system \citep{Cieza16}. In that system the protostar is undergoing a burst of accretion which results in an increased stellar intensity, moving the water ice line far out in the disk ($\sim 42$ AU). This discovery is the first of its kind, where the change of the optical depth of the disk has been directly observed. The water ice line is typically at radii less than $\sim 5$ AU and hence within the resolution limit of our current telescopes. The V883 Ori system is an interesting case study because of its accretion history. As has been discussed, the location and evolution of planet traps (like the ice line) shape the migration and accretion history of protoplanets, as well as impact the radial structure and evolution of the dust. The implications of a stochastic accretion history is beyond the scope of this work.

\cite{And16} points out that the effect of the water ice line should be universal as we expect that every protoplanetary disk will have an ice line. While this is true, we note that this effect evolves in time, and will look different in systems that have drastically different ages. Conveniently, the two protoplanetary disks that have been observed at the highest resolution: HL Tau \citep{ALMA15} and TW Hya \citep{And16} represent systems at either end of the disk evolution. HL Tau is a young ($\sim 1$ Myr) system while TW Hya is an old ($\sim 10$ Myr) one. The closest dark band in HL Tau is at 13.2 AU, while the inner gap in TW Hya is at approximately 1 AU. The location of the dark band in HL Tau has been linked to the water ice line (ie. \cite{Z15}), while \cite{And16} notes that their estimated ice line is within the location of the edge of the TW Hya gap. In Figures \ref{fig:res01a}-\ref{fig:res01c} and \ref{fig:res02a}-\ref{fig:res02c} we see that the water ice line in our model evolves from $\sim 
3.5 $ AU down to $\sim 0.4$ AU which is consistent with the estimates of \cite{And16}. At the very late stages of disk evolution ($>4$ Myr), the disk is primarily heated through direct irradiation from the host star, which produces a temperature profile that does not evolve in time. Hence the location of the ice line will also cease to evolve.

\subsection{ Conclusions }\label{sec:conc}

In this work we have shown that the dust distribution impacts the ionization in the disk. We analyzed the results from two different models of dust evolution: a constant gas-to-dust ratio and well mixed dust model (CPA16-dust) which sets the dust surface density to be proportional to the gas surface density. And a numerical model of dust evolution (Two-pop-dust) where coagulation, fragmentation and radial drift sets the total surface density and size distribution of the dust.
\begin{itemize}
\item Radial drift in the Two-pop-dust clears the dust faster than in the  CPA16-dust model causing a higher ionization
\item In the Two-pop-dust the dead zone edge reduces its physical scale by an order of magnitude when compared to the CPA16-dust model
\begin{itemize}
\item Two-pop-dust: $R_{dz}(t = 0.1$ Myr$) \sim 24$ AU, $R_{dz}(t = 3.7$ Myr$) \sim 0.4$ AU
\item CPA16-dust: $R_{dz}(t = 0.1$ Myr$) \sim 21$ AU, $R_{dz}(t = 3.7$ Myr$) \sim 4.5$ AU
\end{itemize}
\end{itemize}

\noindent A faster evolving dead zone changes the location of the forming protoplanet that is trapped at its edge
\begin{itemize}
\item The relative location of traps with respect to the location of the dead zone impacts whether the planet trap will actually trap a planet. 
\item In the Two-pop-dust model, while the dead zone begins at a higher radii than the heat transition it rapidly evolves within the location of the heat transition trap
\item A planet forming in the heat transition trap will remain trapped over its entire formation in this model
\end{itemize}

\noindent The link between the disk chemical structure and dust physics impacts the retention of the dust 
\begin{itemize}
\item Within the ice line, dust grains are weaker and the maximum grain size is reduced by two orders of magnitude
\item Smaller grains are less susceptible to radial drift and an enhancement of dust forms within the ice line
\end{itemize}

\noindent On top of the rapid radial evolution, the height of the dead zone evolves from $\sim 3$H down to $\sim 1$H near the inner edge of the disk over 3.7 Myr of evolution in the Two-pop-model, while it evolves from $\sim 3$H down to $\sim 2.5$H at the same radial position over the same time in the CPA16-dust model. 
\begin{itemize}
\item Dust settling is less efficient within the ice line, hence the dead zone extends higher within the dead zone, by 2 gas scale heights
\end{itemize}

In a future paper we will connect this model of dust evolution with a full core accretion model of planet formation. This will allow us to directly observe the impact on planet formation caused by this more complicated treatment of dust evolution.

\section*{ Acknowledgements }

We thank our anonymous referee for useful comments that helped to improve the paper. We are grateful to Ted Bergin for the use of his non-equilibrium code. We benefited from many discussions with him and his (former) graduate student Ilse Cleeves on astrochemical codes and analysis. We thank Dmitry Semenov and Cornelis Dullemond for their input during very useful discussions. The work made use of the Shared Hierarchical Academic Research Computating Network (SHARCNET: www.sharcnet.ca) and Compute/Calcul Canada. A.J.C. acknowledges funding from the National Sciences and engineering Research Council (NSERC) through the Alexander Graham Bell CGS/PGS Doctoral Scholarship. R.E.P. is supported by an NSERC Discovery Grant. T.B. acknowledges support from the DFG through SPP 1833 ``Building a Habitable Earth" (KL 1469/13-1). R.E.P. also thanks the MPIA and the Institut f\"ur Theoretische Astrophysik (ITA) in the Zentrum f\"ur Astronomie Heidelberg for support during his sabbatical leave (2015/16) during the final stages of this project. A.J.C also thanks MPIA and the Institut f\"ur Theoretische 
Astrophysik (ITA) in the Zentrum f\"ur Astronomie Heidelberg for their hospitality during his 1 month stay in 2016.

\section*{Appendix}
\appendix
\section{ Testing Further Dust Physics }\label{sec:append01}

Mentioned in the main text is the importance of accurately modeling the dust physics on the resulting ionization structure. Two issues of concern are: 1) properly modeling the vertical extent of the dust and 2) the impact of the location of the dead zone on the dust physics.

We implemented the vertical distribution of \cite{Fro09} and compared the resulting midplane dead zone location with the results of the fiducial model presented in this work - we called this test the FN vert-dist model below. Additionally we explored the impact of an evolving dead zone on the evolution of the dust, while using the same vertical distribution of the fiducial model. Because the turbulence impacts the growth rate, fragmentation and radial distribution of grains, we would expect that implementing a dead zone into the numerical work of \cite{B12} would change how the grains evolves.

The evolving dead zone location defined a radial dependence in the turbulent $\alpha$ parameter that evolves with time. It is modelled by: \begin{align}
\log_{10}(\alpha)(r,t) = -4 + 2\arctan(15\cdot(r - r_{dz}(t)))/\pi
\label{eq:app01}
\end{align}
where $r_{dz}$ is the radial location of the dead zone, and the limits of the function are $-5$ and $-3$ representing the dead zone and active zone respectively. The width of the function, was chosen to match the width used in the fitting of the ice line evolution.

To estimate the location of the dead zone ($r_{dz}$) we fit the evolution of the dead zone location from Figure \ref{fig:dis01b} as a power law of dust surface density, gas surface density and temperature. The fitting function has the form: \begin{align}
r_{dz} = a\Sigma_{dust}^b\Sigma_{gas}^c T^d
\label{eq:app02}
\end{align}
where $a$,$b$,$c$ and $d$ are fitting constants. Because the surface densities and temperature evolve in time, at each time step of the dust evolution model Equation \ref{eq:app02} is solved such that:\begin{align} 
r_{dz} = a\Sigma_{dust}^b(r=r_{dz})\Sigma_{gas}^c(r=r_{dz})T^d(r=r_{dz})
\label{eq:app03}
\end{align}

\begin{figure}
\includegraphics[width=0.5\textwidth]{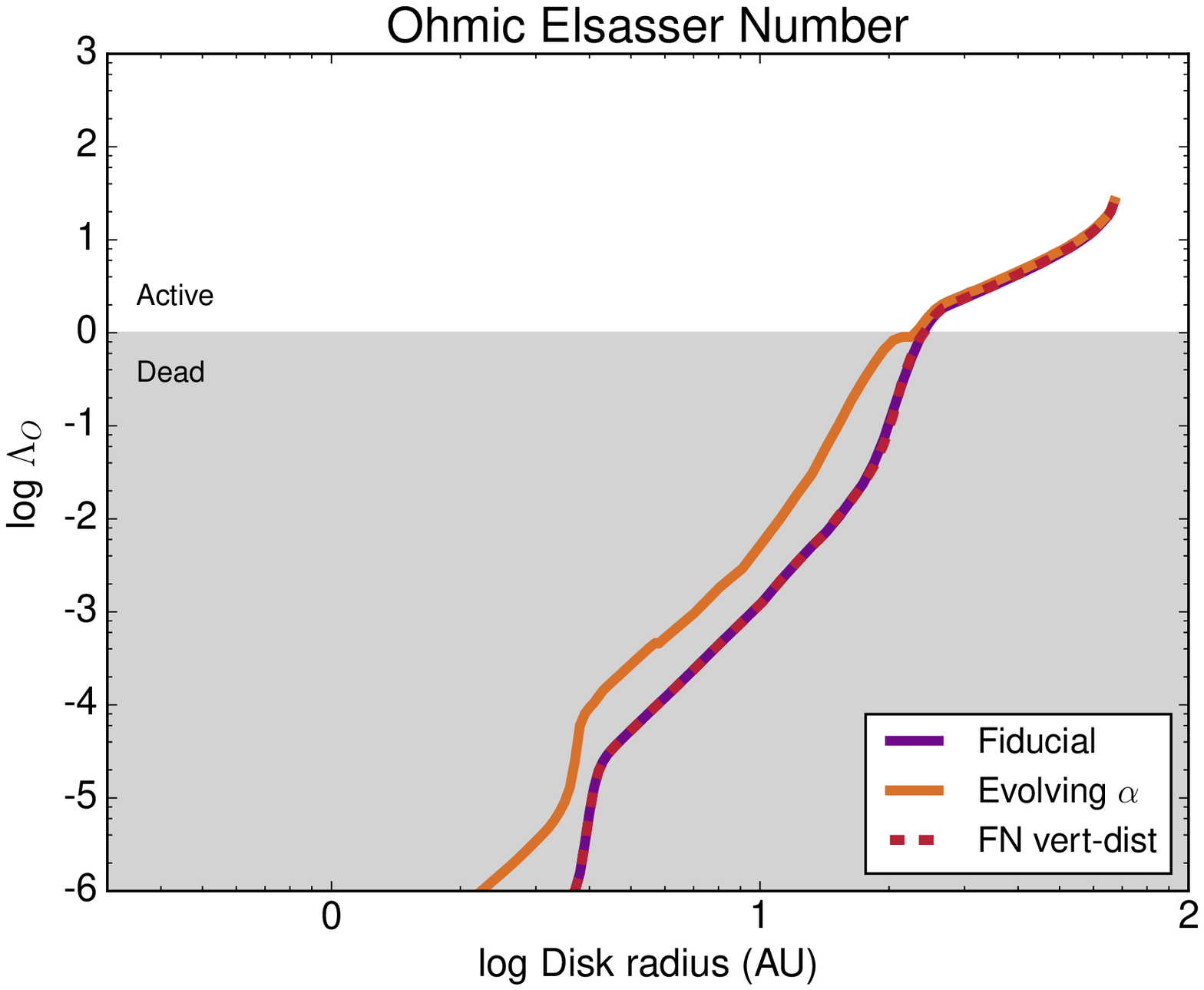}
\caption{The midplane Ohmic Elsasser number for a modified version of the dust vertical distribution (FN vert-dist) and a simple model of the evolving turbulent paramter (Evolving $\alpha$). The FN vert-dist model is indistinguishable from the fiducial model presented in this work.}
\label{fig:app01}
\end{figure}

In Figure \ref{fig:app01} we show the results of testing the new vertical distribution (FN vert-dist) as well as the evolving dead zone location. We find that there is no difference in the location of the midplane dead zone edge, and only small changes in the radial distribution of the Ohmic Elsasser number. In the context of planet formation, these small differences will not result in large changes to the accretion history of forming planets because we assume that the planet is located on the midplane, at the edge of the dead zone.

\begin{figure}
\includegraphics[width=0.5\textwidth]{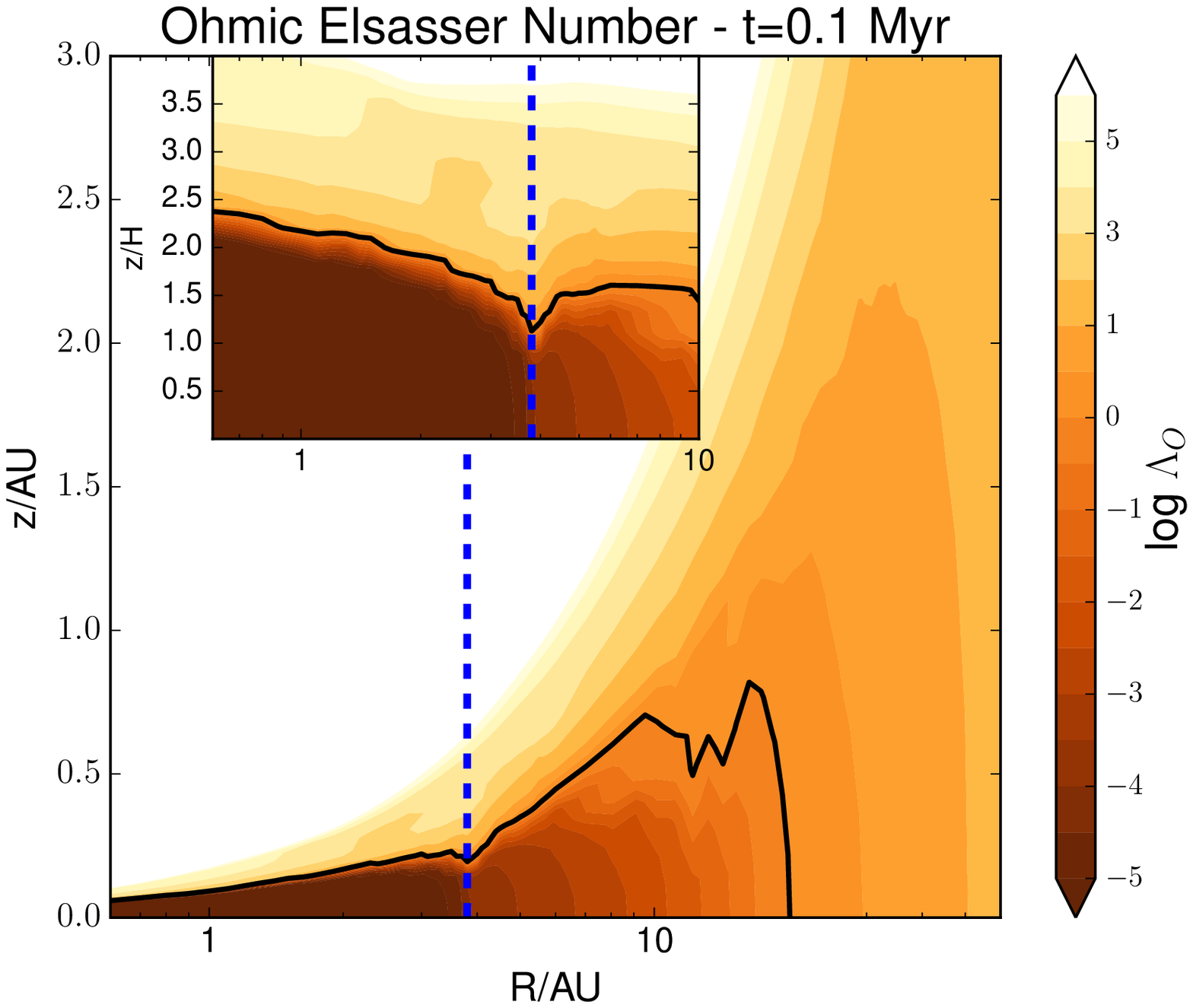}
\caption{ A recreation of Figure \ref{fig:res06a} for the model which includes an evolving radial distribution of the turbulent paramter. With the inclusion of the dead zone, settling is more efficient within the dead zone, and hence the dead zone does not reach as high when compared to the fiducial model. }
\label{fig:app02}
\end{figure}

In Figure \ref{fig:app02} we show the two dimensional map of the Ohmic Elsasser number for the evolving turbulent parameter model. Because the $\alpha$ parameter is lower within the dead zone, settling is more efficient and hence the dead zone does not reach as high in the disk as it did in the fiducial model. Additionally the largest grains are larger within the ice line in this model because fragmentation is generally less efficient within the dead zone. Hence the drastic drop in dead zone height at the radial location of the ice line is not as drastic as it was in the fiducial model.

\begin{figure}
\centering
\subfigure[Dust surface density for the fiducial model. Recreated Figure \ref{fig:res01a}. ]{
	\includegraphics[width=0.5\textwidth]{sig_two_dust_0.1myr.eps}
	\label{fig:app03a}
	}
\subfigure[Dust surface density for the dust model with the evolving dead zone (shown by the grey dotted line).]{
	\includegraphics[width=0.5\textwidth]{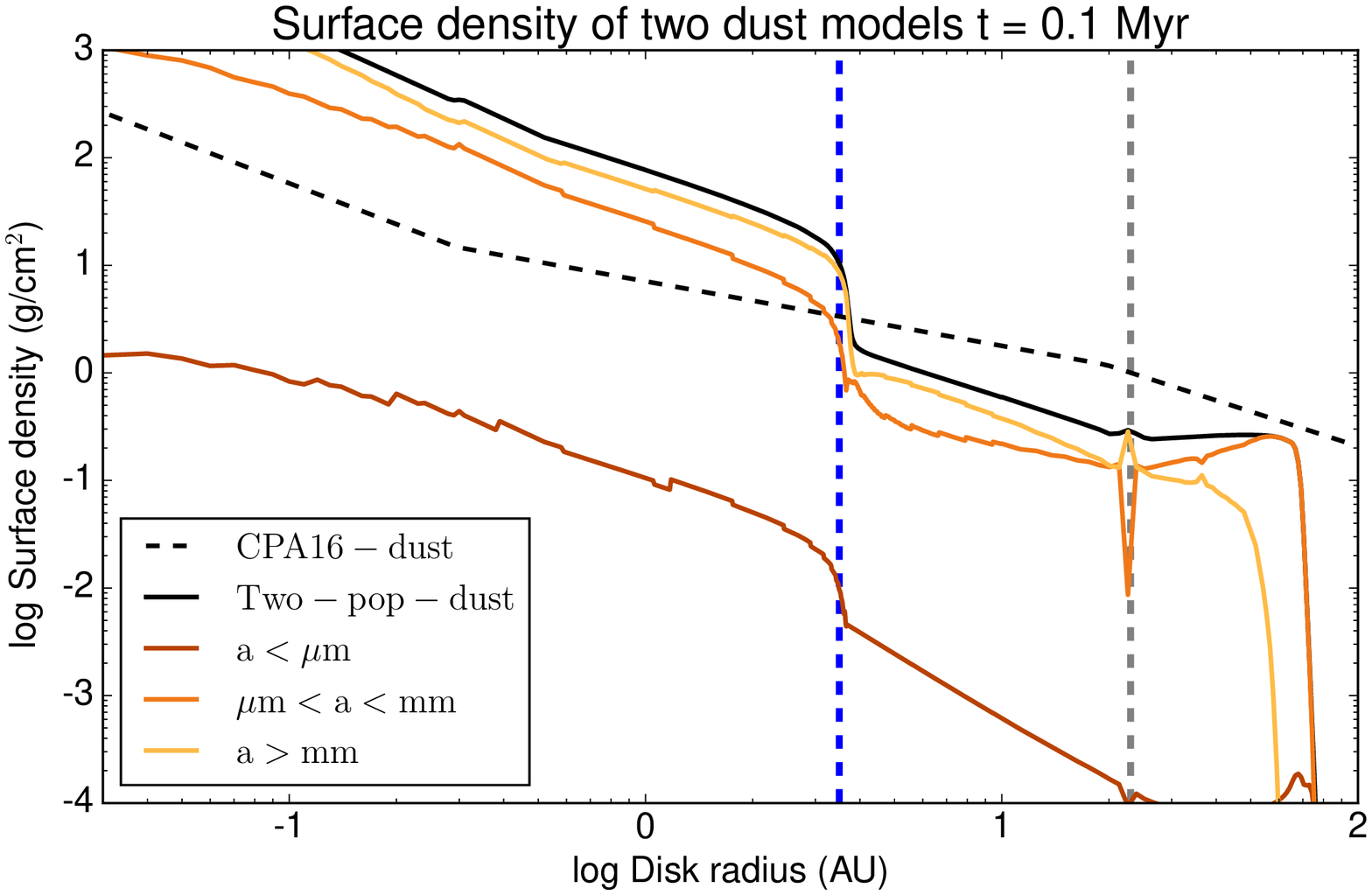}
	\label{fig:app03b}
	}
\end{figure}

In Figures \ref{fig:app03a} and \ref{fig:app03b} we show the dust surface density as a function of grain sizes and radius for the fiducial model, and the model with an evolving dead zone. Importantly we see the impact of reduced fragmentation rate when comparing the surface densities of the sub-micron grains. In the evolving dead zone model, there are less small grains compared to the fiducial model within the ice line, meaning that fragmentation is less efficient within the dead zone. Additonally we find more large grains at the mideplane in the evolving dead zone model because settling is more efficient. These changes in the surface density of different grain sizes could change the long term evolution of the disk because larger grains are more sensitive to radial drift and hence are not retained as long as smaller grains. We explore the long term impact of the evolving dead zone in a future paper.

We find that adding a non-Gaussian vertical distributions for the dust grains shows virtually no change to the distribution of ions that result from the radiative transfer and chemical codes. We find that estimating the location of the dead zone in the dust evolution based on a fitted result of our fiducial model does not change the location of the dead zone edge along the midplane. Hence in the context of planet formation these additional complications do not drastically alter the migration history of a forming protoplanet.

\label{lastpage}

\end{document}